\documentclass[a4]{amfstyle}
\usepackage{times}
\usepackage{amsfonts}
\usepackage{graphicx}
\begin{document}
\title[Nilpotent quantum mechanics, qubits, and flavors of entanglement]
{Nilpotent quantum mechanics, qubits, and flavors of entanglement}
\author[Andrzej M. Frydryszak]{Andrzej M. Frydryszak}
\affiliation{Institute of Theoretical Physics, University of Wroc{\l}aw,
pl. Borna 9, 50-204 Wroc{\l}aw, Poland}
\label{firstpage}
\date{\today}
\maketitle
\begin{abstract}{nilpotent commuting variables, nilpotent mechanics, entanglement, entanglement monotones, Stirling numbers, Bell numbers, symmetric polynomials, supersymmetic quantum mechanics, qubit-fermion systems}
We address the question of description of qubit system in a formalism based on
the nilpotent commuting $\eta$ variables.  In
this formalism qubits exhibit properties of composite
objects being subject of the Pauli exclusion principle, but otherwise behaving
boson-like. They are not fundamental particles. In such an approach the classical limit yields the nilpotent mechanics.

Using the space of $\eta$-wavefunctions, generalized Schr\"{o}\-din\-ger equation etc. we study properties of pure qubit systems and also properties of some composed, hybrid models:
fermion-qubit, boson-qubit. The fermion-qubit system can be truly supersymmetric, with both SUSY partners having identical spectra. It is new and very interesting that SUSY transformations relate here only nilpotent object. The $\eta$-eigenfunctions for the qubit-qubit system give the set of Bloch vectors as a natural basis.

Then the $\eta$-formalism is applied to the description of the pure state entanglement.
Nilpotent commuting variables were firstly used in this context in
[A. Mandilara, et. al., Phys. Rev. A \textbf{74}, 022331 (2006)], we generalize and extend approach presented there.
Our main tool for study the entanglement or separability of states are Wronskians
of $\eta$-functions. The known invariants and entanglement monotones for systems of $n=2,3,4$ qubits are expressed in terms of the Wronskians. This approach gives criteria for separability of states and insight into the flavor of entanglement of the system and simplifies description.
\end{abstract}
\maketitle
\newpage
{\small\em \tableofcontents}
%
\section{Introduction}
Qubit systems \cite{QI} are conventionally described in symmetric tensor products of the two dimensional complex Hilbert spaces. In such an approach there is no classical limit for the qubit understood as a particle. On the other hand, from the supersymetry theory we know that one can consider classical limit for fermions, at cost of introducing the new anticommuting variables. It turns out that similar goal can be achieved for qubits, but this time we have to introduce the nilpotent commuting variables. Such a new formalism is complementary to the conventional one, and gives natural setting to answer the entanglement questions.
The aim of this work is to develop relevant formalism and apply it to the description of entanglement.

As we already mentioned, one-particle space of the qubit is a two dimensional Hilbert space and its collective behavior is a boson-like. In terms of the
Fock space it means that tensor product of qubit states is symmetric. Qubit from this point of view exhibits mixture of the boson and fermion properties. Only separated qubit is like a fermion. To adequately describe qubits we have to play  with the commutation or anticommutation relations on the one side, and
symmetry properties of the tensor product on the other side.  Bosons and fermions can be organized in unique graded structure where commutators, anticommutators, parity of
elements of graded algebra, and symmetry of tensor product is
consistent. They play distinguished role, because they describe fundamental particles. However, there are other useful objects related to parastatistics. Parafermions and parabosons were defined by tri-linear relations in ref. \cite{gren} over fifty years ago and have their place in
quantum field theory \cite{ohka}. Here in our introductory discussion we use the reduced
bi-linear form of these relations, that is satisfactory for our
purposes, but less general then tree-linear relations. Such
reduced form is used in the entries of the Table \ref{CR} concerning
parafermions and parabosons.
A nice discussion of the definition of the qubit, stressing that qubits are
neither bosons nor fermions, is given by  Wu and Lidar \cite{wu-li}.
We departure from their definition of qubit as a parafermion, but find it more
useful to reformulate defining conditions using the commutator and nilpotency conditions. We collect defining properties of relevant objects in the Table \ref{CR} to compare various approaches with canonical commutation relations (CCR), canonical anticommutation relations (CAR), and parabosonic or parafermionic relations.
%
\begin{table}
\caption{Generalized commutation relations.}\label{CR}
\begin{center}
\begin{tabular}{lrrc}
\hline\\[-5mm]
\hline\\[-2mm]
Type & Single site rel. & Different site rel. & Symmetry\\[0.5mm]
 \hline\\[1mm]
&&&\\
Bosons & $[b,b^+]_-=1 $ & $[b_i, b^+_j]_-=0 $ & +\\
(CCR algebra)     &$[b,b]_-=0 $ & $[b_i, b_j]_-=0$&\\
       &$[b^+,b^+]_-=0$ & $[b^+_i, b^+_j]_-=0$&\\
&&&\\[-2mm]
\hline\\
Fermions&$[f,f^+]_+=1 $ & $[f_i, f^+_j]_+=0 $ & --\\
(CAR algebra)&$[f,f]_+=0 $ & $[f_i, f_j]_+=0$&\\
       &$[f^+,f^+]_+=0$ & $[f^+_i, f^+_j]_+=0$&\\
&&&\\[-2mm]
\hline\\
Parabosons & $[e,e^+]_-=1-2 \lambda K   $ & $[e_i, e^+_j]_-=0 $ & +\\
 &$[e,e]_-=0 $ & $[e_i, e_j]_-=0$&\\
       &$[e^+,e^+]_-=0$ & $[e^+_i, e^+_j]_-=0$&\\
&&&\\[-2mm]
\hline\\
Parafermions&$[a,a^+]_+=1 $ & $[a_i, a^+_j]_-=0 $ & +\\
(Spin algebra)&$[a,a]_+=0 $ & $[a_i, a_j]_-=0$&\\
       &$[a^+,a^+]_+=0$ & $[a^+_i, a^+_j]_-=0$&\\
&&&\\[-2mm]
\hline\\
Qubits&$[d,d^+]_-=1-2N $ & $[d_i, d^+_j]_-=0 $ & +\\
(Qubit algebra)&$d^2=0 $ & $[d_i, d_j]_-=0$&\\
       &$(d^+)^2=0$ & $[d^+_i, d^+_j]_-=0$&\\[2mm]
\hline\\[-5mm]
\hline\\[-3mm]
\end{tabular}
\end{center}
\end{table}
%
In the last column there are given  eigenvalues $\pm 1$ of the
transposition (flip) map $\tau$, where
$\tau_{ii+1}(...\otimes\psi_i\otimes\psi_{i+1} ...)
= ...\otimes\psi_{i+1}\otimes\psi_i ...$ .
The two dimensional state space of a single qubit system is naturally obtained from
nilpotent  creation/anihilation operators.
Taking commutators (symmetric tensor products) for many-site system we obtain parafermions.
This approach is adopted most frequently cf. Ref.\cite{wu-li}.
To find another description of qubits we rewrite the set of (anti)commutation relations for parafermions in the following way
\begin{eqnarray}
[d,d^+]_-=1-2N, \quad\quad &&[d_i, d^+_j]_-=0, \quad\quad [d_i, d_j]_-=0,
\quad\quad [d^+_i, d^+_j]_-=0,\\\nonumber
&& \quad\quad d^2=0, \quad\quad (d^+)^2=0,
\end{eqnarray}
where $N$ can be seen as particle number operator.
As it is known there is no classical limit, in the usual sense, for fermions. Only introduction of the Grassmannian variables allows to define such a limit. Analogously for qubits, to get nontrivial classical limit we have to introduce nilpotent, but this time - commuting variables.
Namely,
\begin{equation}
\eta \eta'=\eta'\eta, \quad \quad \eta^2=\eta'^2=0
\end{equation}
This provides that single qubit is two-level one, in the same time it is boson-like
when considered in multi-qubit system.
In the approach, where the qubits are realized within the conventionally defined
spin algebra  (parafermions),
natural classical limit of such system yields the
anticommuting co-ordinates \cite{wu-li}, like for fermions. Such anticommuting coordinates imply
naturally the antisymmetry of the tensor product, and nilpotent "bosonic" objects again have to be represented by even product of anticommuting variables. That is why parafermions in such a setting do not suit our demands. To describe qubit without referring to its composite character one has to use nilpotent commuting variables.
But we have to pay for using the nilpotent $\eta$-variables; the
$\eta$-"derivative" does not fulfil the Leibniz rule \cite{amf-ncm, fik}. On the
other hand, the system of many qubits in the $\eta$-formalism has
the same property as observed  in \cite{wu-li} for parafermions, that it behaves for large $n$ like boson, in paricular becames not nilpotent.
Already in the 1994 Palumbo \cite{palu-4} noted, in the context of high energy
physics, that nilpotent commuting variables:
\begin{quotation}
[nilpotent commuting variables] are not just abstract mathematical
entities devoid of any physical interpretation, but describe composites of
fermions, so that models can be of phenomenological importance.
\end{quotation}
\par
The formalism of
$\eta$-functions which later will be used to describe the nilpotent
quantum mechanics was discussed in more detail in \cite{amf-ncm, amf-n1, amf-nf}. Properties of such functions allow to answer many questions related to the entanglement. Criteria of factorization of states coming from this formalism are natural, are derived by simple argument and are equivalent to the ones known from the invariants theory approach. In the work of Wu and Lidar there is  analysis of the notion of qubit from the particle point of view and is given the parafermionic description of this object. In our approach we rewrite the parafermionic character of qubit in such a way that the commutators and nilpotency conditions are used in defining relations, but resulting properties are analogous as in \cite{wu-li}. Despite presented above fundamental arguments for representing qubits with use of the $\eta$-variables, arguments related to symmetry of qubit state product and nilpotency; it was found by Mandilara et al. in \cite{mandi, mandi2} that nilpotent commuting variables are very useful tool in the description of entanglement. One may think that such formalism has deeper roots, and like for fermions one can consider (pseudo)classical systems of superparticles, then quantize them. Here we should have the formalism of nilpotent mechanics for systems that after quantization will give the (quantum) qubit systems.

Nilpotent commuting variables, besides the quantum mechanics \cite{mandi}, are present in the theoretical physics in several contexts: nuclear and high energy physics, string theory \cite{palu-1, palu-2, palu-3, palu-4, palu-5}, as well as in metioned before classical nilpotent mechanics. For review of some applications cf. Ref. \cite{amf-nf}.

The paper is organized as follows. In the next section we
describe the qubit system in such a way that it has the proposed above classical limit realized by the commuting nilpotent coordinates $\eta$. Then necessary formalism based on $\eta$-coordinates is presented. We recall definition of
the $\eta$-numbers, $\eta$-functions \cite{amf-ncm} and then develop it by studying properties of the elementary $\eta$-functions, in particular:  exponent, logarithm, and trigonometric functions. It turns out that the Stirling numbers, the ordered Bell numbers, and the second kind Bell numbers are related to the expansions of specific $\eta$-functions. The following section we devote to the introduction of the symmetric $\eta$-polynomials and Hermite $\eta$-polynomials. Then the rudiments of the $\eta$-calculus are given with the $\eta$-integral, the $\eta$-Fourier transform as well as the $\eta$-integral form of the Stirling and the normal Bell numbers. Having above tools we introduce the notion of the $\mathcal{N}$-Hilbert space which will allow to describe qubit systems. Using the $\eta$-integral we define the $\mathcal{N}$-scalar product in the space of the $\eta$-functions (wave $\eta$-functions) and the structure of the $\mathcal{N}$-Hilbert space. It turns out that natural bases in such spaces are closely related to the ones used in the description of the entanglement of qubit systems. Here, such bases have natural geometrical meaning. We also present $\eta$-kernels of some basic operators acting in $\mathcal{N}$-Hilbert space.

In the $\mathcal{N}$-Hilbert space of the wave $\eta$-functions one can consider the generalized Schr\"{o}\-din\-ger equation. Formerly, such equation, but in a different form (logarithmic form), was proposed by Mandiliara et al. \cite{mandi}. In the Section \ref{2comp} we study twocomponent nilpotent systems and the eigenvalue problem for qubit system covering various types of couplings (Ising, XY etc.). It is interesting that the set of eigenvector for these systems is composed of the Bell vectors. There are also considered composed systems of the  qubit and boson, the qubit and fermion, and the qubit-qubit. As specially interesting we find the qubit-fermion system with the supersymmetry acting between two nilpotent parts.

In the Sec. \ref{sep} there is addressed question of the separability of the  $\eta$-variable dependence of the $\eta$-functions and a "duality" of such notion to the entanglement. In the conventional function theory, the separability questions are answered using the appropriate Wronskians \cite{simsa}.
Here we find that there can be considered special generalizations of Wronski matrices, and that their traces and determinants serve as a tool to detect the conditions for separability of functions. What is remarkable, one can express relevant invariants known from the general theory and entanglement monotones used in recent works on entanglement, in terms of Wronskians. This gives additional insight into the structure or flavor of entanglement. In the description we follow the number of qubits, $n=2,3,4$. In the present work we answer only the question
of the pure state entanglement. Finally, conclusions are given in Sec.\ref{concl}. The work is accompanied with several Appendices containing relevant auxiliary material and derivations of important facts.
 %
%
\section{Canonical qubit relations}\label{QR}
As we have mentioned in the introduction we shall follow  Wu and Lidar, and by
the qubit we denote  a two-level quantum object with encoded boson-like behavior in many particle system, and in the same time, underlying the Pauli exclusion principle. As it was already noted in Ref. \cite{palu-1, palu-2, palu-3, palu-5, wu-li, mandi, amf-n1, amf-ncm,  amf-nf} such a somehow hybrid object is not a fundamental particle, like boson or fermion and inherently carries properties of a composed object, but it can be described without any explicit reference to the constituents and the way it is composed of. The formalism we develop provides independent description of the qubit itself and gives very convenient characterization of qubit pure state entanglement. The nilpotent commuting variables were already used in the context of entanglement by Maniliara at al. in \cite{mandi}.%
\par
In the analogous way as the fermion and fermionic states are
understood, one can define the qubit system  by means of the so
called linear commutation relations to have on the same footing
canonical commutation relations (CCR), canonical anticommutation
relations (CAR) and canonical qubit relations (CQR). To take into account boson-like behavior of  qubit we prefer to use only commutators, and realize the Pauli exclusion principle, imposing nilpotency conditions without referring to anticommutators. Such defined object is parafermionic, but the form of relations stresses its boson-like properties. This is also important from the point of view of supersymmetry. As we shall see later one can define supersymmetric system including qubits. Within present scheme, qubits will be even object in the graded (supersymmetric) structure, and underly commutation relations as such.

To describe qubit as an object which is nilpotent but otherwise
boson-like, let us observe that the commutator of qubit creation
operator $d^+$ and qubit anihilation operator $d$ if nontrivial,
cannot have a value in the center of
 commutator algebra. This would give an contradiction with nilpotency. Therefore
let us make the following ansatz
\begin{equation}
[\,d, d^+]_- =1-2N
\end{equation}
On the other hand compatibility condition with the nilpotency of $d$, $d^+$ takes the following form
\begin{equation}
[\,d, d^+]_+ = 1+2Z,
\end{equation}
where $Z=(d^+d-N)$ is an element from the center of the algebra. Moreover
\begin{equation}
[ N, d\,\,]_+  = d
\end{equation}
\begin{equation}
[ N, d^+]_+ = d^+
\end{equation}
The canonical qubit relations (CQR) consist of the
following set of conditions
\begin{eqnarray}\label{cqr1}
&&\left[\,d, d^+ \right]_- = 1-2N\\
&&\left[N, d\,\,\right]_- = d\\
&&\left[N, d^+ \right]_- = -d^+\\
&&d^2=(d^+)^2 = 0 \label{cqr4}
\end{eqnarray}
In particular one can take $N=d^+d$.
For fermionic operators we have $(a+a^+)^2=1$, and here for
qubit operators
\begin{equation}
(d+d^+)^2=1+2Z,
\end{equation}
where for the $N=d^+d$,  $Z=0$. As we shall show in the following sections, the CQR can be naturally realized within the $\eta$-function space by means of $\eta$-differential and multiplication operators.
\par
Using for the CQR the argument of Wu and Lidar
(\cite{wu-li}) one can see that the set of $n$ qubits for the large
$n$ behaves like a boson. Namely
\begin{eqnarray}
b&=&\frac{1}{\sqrt{n}}\sum_{i=1}^{n}d_i\\
b^+&=&\frac{1}{\sqrt{n}}\sum_{i=1}^{n}d^+_i
\end{eqnarray}
then
\begin{equation}
[b, b^+]_-=1-\frac{2}{n}\sum_{i=1}^n N_i.
\end{equation}
Hence, when the number of sites $n$ is much larger then the
effective number of qubits we get $[b, b^+]_-\approx 1$.
%
%
\section{$\eta$-formalism}
We want to realize the CQR relations in the space of functions of the commuting nilpotent variables $\eta$, what suggests the classical limit considered in the Introduction. Because of the nilpotency of the $\eta$-variables we can expect some similarities to the superanalysis, but this two formalisms are different. The nilpotency is natural for anticommuting (odd) elements but for commuting (even) elements it is a result of restrictive condition. In the following sections let us collect and introduce some necessary notions.
%
\subsection{$\mathcal{N}$-numbers}
To introduce necessary functions for describing system of qubits we shall consider nilpotent commuting $\eta$-variables defined in Ref. \cite{amf-ncm}
as special elements of an algebra $\mathcal{N}$.
We recall, that the algebra $\mathcal{N}$ is freely generated by the set of first order nilpotents and the unit. Each element $\nu \in \mathcal{N}$ can be decomposed into the numerical (real or complex) part called body and into the nilpotent part called soul
\begin{equation}
\nu= {b}(\nu)+{s}(\nu)
\end{equation}
Any element with a nonzero body has an inverse
\begin{equation}
\nu^{-1}=b(\nu)^{-1}\sum_{m=0}^{\infty}(-b(\nu)^{-1}s(\nu))^m
\end{equation}
Analogous formula is known for superalgebra \cite{dewitt, rogers}. It is useful to note, that for elements with $b(\nu)=1$ it takes the form
\begin{equation}
(1+s(\nu))^{-1}=\sum_{m=0}^{\infty}(-s(\nu))^m
\end{equation}
literary generalizing the celebrated formula for real numbers $\frac{1}{1+t}=\sum(-t)^m$.

The $\eta$-variables are first order nilpotents, $\eta\in \mathcal{N}$, $\eta^2=0$. They form subset $\mathcal{D}\subset\mathcal{N}$. Let $\eta\neq\eta'$ and $t\in\mathbb{R}$, $0\leq t\leq 1$ then
\begin{equation}
(t\eta+(1-t)\eta)^2=2t(t-1)\eta\eta'
\end{equation}
Therefore $\eta$'s algebraically dependent with fixed $\eta$ element (i.e. $\eta\eta'=0$) form a star-convex set. For algebraically independent elements
it is not true unless we admit other, higher order nilpotents to enter the set. Further details on the $\mathcal{N}$ algebra can be found in \cite{amf-ncm}.
%
\subsection{$\eta$-functions}
One can define functions depending on $\eta$-variables, where these variables are not fixed "generators" but can vary within some set \cite{amf-ncm}. If we take
$\eta_i$, $i=1,2,\dots, n$ to be a set of independent nilpotent elements i.e.
\begin{equation}
\eta_i^2=0 \;\; \forall \,  i, \quad \eta_1\cdot \eta_2 \cdot \dots \eta_n\neq 0
\end{equation}
and $\vec{\eta}=(\eta_1, \eta_2, \dots \eta_n)$. Using notation
that $I_0=\emptyset$ and $I_k=({i_1}, {i_2}, \dots {i_k})$ is ordered multi-index. One can write the expansion of function $F(x, \vec{\eta})\in \mathcal{F}[\vec{\eta}]$ of the $n$
$\eta$-variables in the general form
\begin{equation}
F(x,\vec{\eta})=\sum_{k, I_k}^n F_{I_k}(x)\eta_{I_k},
\end{equation}
where $F_{I_k}(x)\in \mathcal{N}$ and $\eta_{I_0}=1$.
In the present paper we shall consider only functions with
$F_{I_k}(x)\in \mathbb{R}, \mathbb{C}$. This set we shall denote by $\mathcal{F}_0[x,\vec{\eta}]$. Let us observe that there is a natural mapping $\Xi$ from the set of symmetric $n\times n$ matrices to the $\mathcal{F}_0[\eta^1,\dots,\eta^n]$. We take as components $F_{I_k}$ the principal minors of a matrix with entries indexed by $I_k=(i_1, i_2,\dots,i_k)$, moreover we assume that  $F_{\emptyset}=1$ e.g.
\begin{equation}
B=\left(
\begin{array}{ll}
b_{11}&b_{12}\\
b_{21}&b_{22}
\end{array}
\right)\mapsto\Xi(B)=
F(\eta^1,\eta^2)=1+b_{11}\eta^1+b_{22}\eta^2+(b_{11}b_{22}-b_{12}^2)\eta^1\eta^2
\end{equation}
In the set of $\eta$-functions $\mathcal{F}_n(\vec{\eta})$
there exists $\mathbb{Z}_2$ gradation related to the decomposition into
the sets of even and odd functions. The gradation mapping $J$ can be
defined as
usual by relations
\begin{equation}
J(1)=1, \quad J(\eta_i)=-\eta_i, \mbox{and}\quad J(\eta^{I_k})=J{(\eta^{i_1})J(\eta^{i_2})\dots J(\eta^{i_k})}
\end{equation}
Another important operation which will be frequently used is duality transformation. It resembles the the Hodge $\star$ - operator known for exterior forms. The duality operator $\Theta$ we shall define as follows
\begin{equation}
\Theta(\eta_{I_k})=\eta_{I_{n-k}}, \quad\quad I_k\cup I_{n-k}=I_n,
\end{equation}
naturally $\Theta(1)=\eta_1\eta_2\dots\eta_n$ and $\Theta^2=id$.
Functions such, that $\Theta(F(\vec{\eta}))=F(\vec{\eta})$ we shall call selfdual, and these with $\Theta(F(\vec{\eta}))=-F(\vec{\eta})$ antiselfdual. In the sequel we shall use $\Theta$ notation for this operation or we will denote dualization by $\star$, i.e. $\star F\equiv\Theta(F)$.

In the next section, in view of further applications we consider some elementary $\eta$-functions.
\subsubsection{Elementary $\eta$-functions}
To describe qubit systems we shall use elementary $\eta$-functions. They can be defined by means of a series
analogous to the conventional elementary functions. Firstly we introduce explicit form of powers of the $\mathcal{F}_0[x,\vec{\eta}]$ for $n=1,2,3,4$. In the following we shall omit the $x$-variable dependence. Providing explicit formulas we have in mind further concrete considerations for $n$-qubit systems, $n=1,2,3,4$.
\begin{itemize}
\item[] \underline{\bf Power function.}
In the formulas below we sum over strictly ordered configurations of indices and appropriate multiplicities of terms are taken into account.\\
\centerline{${\bf n=1}$:}
\begin{equation}
F(\eta)=F_0+F_1\eta
\end{equation}
\begin{equation}
F(\eta)^m=F_0^{m-1}(F_0+F_1\eta),
\end{equation}
\centerline{${\bf n=2}$:}
\begin{equation}
F(\eta^1,\eta^2)=F_0+F_1\eta^1+F_2\eta^2+F_{12}\eta^1\eta^2=F_0+F_i\eta^i+F_{ij}\eta^i\eta^j
\end{equation}
\begin{equation}
F(\eta^1,\eta^2)^m=F_0^{m-1}(F_0+mF^i\eta_i)+F_0^{m-2}(n F_0 F_{12}+m(m-1)F_1F_2)\eta_1\eta_2
\end{equation}
\centerline{${\bf n=3}$:}
\begin{eqnarray}\nonumber
F(\eta^1,\eta^2, \eta^3)
=&&F_0+F_1\eta^1+F_1\eta^2+F_1\eta^3
+F_{12}\eta^1\eta^2+F_{13}\eta^1\eta^3+F_{23}\eta^2\eta^3\\
&&+F_{123}\eta^1\eta^2\eta^3=
F_0+F_i\eta^i+F_{ij}\eta^i\eta^j+F_{ijk}\eta^i\eta^j\eta^k
\end{eqnarray}
\begin{eqnarray}\nonumber
F(\eta^1,\eta^2, \eta^3)^m=&&
F_0^{m-1}(F_0+mF_i\eta^i)+F_0^{m-2}(mF_0F_{12}\\\nonumber
&&+m(m-1)F_1F_2)\eta_1\eta_2
+F_0^{m-2}(mF_0F_{13}\\
&&+m(m-1)F_1F_3)\eta_1\eta_3
+F_0^{m-2}(mF_0F_{23}\\\nonumber
&&+m(m-1)F_2F_3)\eta_2\eta_3+
F_0^{m-3}(m(m-1)F_0(F_1F_{23}\\\nonumber
&&+F_2F_{13}+F_3F_{12})+
+m(m-1)(m-2)F_1F_2F_3\\\nonumber
&&+mF_0^2F_{123})\eta^1\eta^2\eta^3
=F_0^{m-1}(F_0+mF_i\eta^i)\\\nonumber
&&+F_0^{m-2}(mF_0F_{ij}+m(m-1)F_iF_j)\eta_i\eta_j
+F_0^{m-3}(m(m-1)\\\nonumber
&&F_0F_iF_{jk}
+m(m-1)(m-2)F_iF_jF_k
+mF_0^2F_{ijk})\eta^i\eta^j\eta^k
\end{eqnarray}
\centerline{${\bf n=4}$:}
\begin{equation}
F(\eta^1,\eta^2, \eta^3,
\eta^4)=F_0+F_i\eta^i+F_{ij}\eta^i\eta^j+F_{ijk}\eta^i\eta^j\eta^k+F_{ijkl}\eta^i\eta^j\eta^k\eta^l
\end{equation}
\begin{eqnarray}\nonumber
F(\eta^1,\eta^2, \eta^3,
\eta^4)^m=&&F_0^{m-1}(F_0+mF_i\eta^i)+F_0^{m-2}(mF_0F_{ij}+m(m-1)\\\nonumber
&&F_iF_j)\eta_i\eta_j+F_0^{m-3}(mF_0^2F_{ijk}\\\nonumber
&&+m(m-1)F_0F_iF_{jk}+m(m-1)(m-2)\\
&&\cdot F_iF_jF_k)\eta^i\eta^j\eta^k
+F_0^{m-4}(nF_0^3F_{ijkl}\\\nonumber
&&+m(m-1)F_0F_iF_{jkl}
+m(m-1)F_0F_{ij}F_{kl}\\\nonumber
&&+m(m-1)(m-2)(m-3)\cdot F_iF_jF_kF_l)\eta^i\eta^j\eta^k\eta^l
\end{eqnarray}
A cautionary remark: to put the formulas for powers of arbitrary
$F(\vec{\eta})$ in a compact form we use the following conventions: when term
gets negative power of $(F_0)^{m-k}$ - it vanishes; for
the terms with the factor $(F_0)^0$ in front, we put $(F_0)^0=1$ . Only after such
preliminary adjustments we substitute actual values of $F_{I_k}$.
In the present work we will be satisfied with the above formulas for
$n\leq 4$, but generalization to higher $n$ is straightforward.
\item[] \underline{\bf Exponent.}
The $\eta$-exponent we define using conventional expansion, what gives
\begin{equation}
e^{F(\vec{\eta})}=\sum_{n=0}^{\infty}\frac{F(\vec{\eta})^n}{n!}=e^{F_0}e^{s(F(\vec{\eta}))}
\end{equation}
\item[]\underline{\bf Logarithm.} Again definition of logarithm function is conventional but it turns out that its terms in the expansion have very interesting relations to hyperdeterminants. In this context logarithm of functions of commuting nilpotent variables was considered in \cite{mandi}. We shall come back to this in the context of separability.
For $\eta$-functions with unit body one can define
\begin{equation}
ln(1+s(F(\vec{\eta}))=\sum_{k=1}^{\infty}(-1)^{k-1}\frac{s(F(\vec{\eta}))^k}{k},
\end{equation}
where $s(F(\vec{\eta}))$ is the soul of function $F(\vec{\eta})$ i.e. here $F(\vec{\eta})=1+s(F(\vec{\eta}))$. Again let us consider explicit formulas for $n=1, 2, 3, 4$ using expansions with $F_0=1$.
\centerline{${\bf n=1}$:}
\begin{equation}
ln(1+s(F({\eta}))=F_1\eta
\end{equation}
\centerline{${\bf n=2}$:}
\begin{equation}
ln(1+s(F(\eta^1,\eta^2))=F_i\eta^i+(F_{ij}-F_iF_j)\eta^i\eta^j
\end{equation}
\centerline{${\bf n=3}$:}
\begin{eqnarray}
ln(1+s(F(\eta^1,\eta^2,\eta^3))&=&F_i\eta^i+(F_{ij}-F_iF_j)\eta^i\eta^j\\\nonumber
&+&(F_{ijk}-F_iF_{jk}+2F_iF_jF_k)\eta^i\eta^j\eta^k
\end{eqnarray}
Using notation $f(\vec{\eta})=ln(1+s(F({\eta}))=ln(F({\eta}))$ we can find the following identity for components of logarithm for functions belonging to the image of the mapping $\Xi$ for $n=3$ \cite{holtz-stur}. Let us note that this identity was derived in \cite{holtz-stur} without any relation to the $\eta$-logarithm.
\begin{equation}
f^2_{123}=-4f_{12}f_{13}f_{23}=4b_{12}^2b_{13}^2b_{23}^2
\end{equation}
\centerline{${\bf n=4}$:}
\begin{eqnarray}\nonumber
ln(1+s(F(\eta^1,\eta^2,\eta^3,\eta^4))
&=&F_i\eta^i+(F_{ij}-F_iF_j)\eta^i\eta^j+
(F_{ijk}-F_iF_{jk}\\
&+&2F_iF_jF_k)\eta^i\eta^j\eta^k
+(F_{ijkl}-F_{ij}F_{kl}\\\nonumber
&-&F_iF_{jkl}+2F_iF_jF_{kl}
-6F_iF_jF_kF_l)\eta^i\eta^j\eta^k\eta^l
\end{eqnarray}
Here as well there exist identities for components of logarithm of function from the image of the $\Xi$ mapping, namely \cite{holtz-stur}
\begin{equation}
f_{ijk}f_{ijl}f_{ikl}=-4f_{ikl}f_{ij}f_{ij}f_{ik}f_{jl}
\end{equation}
\begin{equation}
-2f_{ijkl}f_{ij}f_{ik}f_{il}=f_{ikl}f_{ijk}f_{ij}f_{il}+f_{ikl}f_{ijl}f_{ij}f_{ik}+
f_{ijl}f_{ijk}f_{ik}f_{il}
\end{equation}
\item[] \underline{\bf Trigonometric functions.} This family of $\eta$-functions is  defined by formal series analogous to the conventional one, namely
let $F=\sum F_{I_k}\eta^{I_k}$, $F_{I_k}\in \mathcal{R}$
\begin{equation}
cos(F(\vec{\eta}))=\sum_{k=0}(-1)^k\frac{F^{2k}}{(2k)!}
\end{equation}
\begin{equation}
sin(F(\vec{\eta}))=\sum_{k=0}(-1)^k\frac{F^{2k+1}}{(2k+1)!}
\end{equation}
Obviously $cos^2(F) + sin^2 (F)=1$ and we have
\begin{eqnarray}
cos(F)&=&cos(F_0)cos(s(F))-sin(F_0)sin(s(F))\\
sin(F)&=&sin(F_0)cos(s(F))+cos(F_0)sin(s(F))
\end{eqnarray}
for a function of one $\eta$ variable we obtain that
\begin{equation}
\left(
\begin{array}{l}
cos(F(\eta))\\
sin(F(\eta))
\end{array}
\right)
=
\left(
\begin{array}{lr}
cos(F_0)&-sin(F_0)\\
sin(F_0)&cos(F_0)
\end{array}
\right)
\left(
\begin{array}{c}
1\\
F_1\eta
\end{array}
\right)
\end{equation}
\end{itemize}
For further considerations we will need $sin(\sum\eta^i)$ and $cos(\sum\eta^i)$ in explicit form for $n=2,3,4$.\\[2mm]
\centerline{${\bf n=2}$:}
\begin{eqnarray}\label{trig-2}
cos(\eta^1+\eta^2)&=&1-\eta^1\eta^2\\
sin(\eta^1+\eta^2)&=&\eta^1+\eta^2
\end{eqnarray}
Let us note that $cos(\eta^1+\eta^2)$ is antiselfdual and $sin(\eta^1+\eta^2)$ is selfdual.\\[5mm]
\centerline{${\bf n=3}$:}
\begin{eqnarray}\label{trig-3}
cos(\eta^1+\eta^2+\eta^3)&=&1-\eta^1\eta^2-\eta^1\eta^3-\eta^2\eta^3\\
sin(\eta^1+\eta^2+\eta^3)&=&\eta^1+\eta^2+\eta^3-\eta^1\eta^2\eta^3
\end{eqnarray}
Here $\Theta(cos(\eta^1+\eta^2+\eta^3))=-sin(\eta^1+\eta^2+\eta^3)$.\\[5mm]
\centerline{${\bf d=4}$:}
\begin{eqnarray}\label{trig-4}
cos(\eta^1+\eta^2+\eta^3+\eta^4)
&=&1-\eta^1\eta^2-\eta^1\eta^3-\eta^1\eta^4-\eta^2\eta^3\\\nonumber
&-&\eta^2\eta^4-\eta^3\eta^4+\eta^1\eta^2\eta^3\eta^4\\
sin(\eta^1+\eta^2+\eta^3+\eta^4)&=&\eta^1+\eta^2+\eta^3+\eta^4
-\eta^1\eta^2\eta^3-\eta^1\eta^2\eta^4\\\nonumber
&-&\eta^1\eta^3\eta^4-\eta^2\eta^3\eta^4
\end{eqnarray}
In this case $cos(\eta^1+\eta^2+\eta^3+\eta^4)$ is selfdual and
$sin(\eta^1+\eta^2+\eta^3+\eta^4)$ is antiselfdual.
%
\subsection{Distinguished $\eta$-functions}
It turns out that $\eta$-functions be related to combinatorics and graph theory.
Here we want to describe a family of $\eta$ functions yielding the normal and ordered Bell numbers and Stirling numbers of the second kind.
\begin{itemize}
\item[]\underline{\bf Functions: $\mathcal{T}_n$ and $\mathcal{E}_n$.}
Let us define the following $\eta$-functions
\begin{equation}
\mathcal{T}_n=\mathcal{T}_n(\eta_1,\eta_2,\dots,\eta_n)=\sum_{i=1}^n\eta^i
\end{equation}
and
\begin{equation}
\mathcal{E}_n=\mathcal{E}_n(\eta_1,\eta_2,\dots,\eta_n)=\sum_{k=1, I_k}^n\eta^{I_k}
\end{equation}
As we shall see they play important role in describing combinatorial characteristic in terms of $\eta$ variables.
Directly from definitions we have the following relations
\begin{equation}
\mathcal{E}_n=e^{\mathcal{T}_n}-1=s(e^{\mathcal{T}_n})
\end{equation}
\end{itemize}
Expanding k$^{th}$ power of $\mathcal{E}_n$ we find that numerical coefficients in respective terms of expansion just count the number of functions from $m$-element set into an $k$-element set, giving the following formula \cite{note1}
\begin{equation}\label{Stirling}
(\mathcal{E}_n)^k=\sum_{k\leq m\leq n}\sum_{I_m}k!\,S(m,k)\,\eta^{I_m},
\end{equation}
where $S(m,k)$ are Stirling numbers of the second kind. Because the following sum gives the Bell number $B_n$ (so called normal Bell number)
\begin{equation}\label{Bell1}
B_n=\sum_{m=0}S(n,m),\quad n\geq 1,
\end{equation}
from the Eq. (\ref{Stirling}) we get that
\begin{equation}\label{Bell2}
e^{\mathcal{E}_n}=\sum_{k=0}\sum_{I_k}\,B_k\,\eta^{I_k},
\end{equation}
It is easy to prove  another important relation which involves ordered Bell numbers $C_n$, namely
\begin{equation}\label{Bell}
(1-\mathcal{E}_n)^{-1}=\sum_{k=0}\sum_{I_k}\,C_k\,\eta^{I_k},
\end{equation}
For convenience of the reader let us recall the values of Bell numbers for $n=0,1,\dots,5$. Namely,  $B_n:\, 1,1,2,5,15,52$ and $C_n:\, 1,1,3,13,75,541$, respectively.
%
\subsection{Symmetric functions}
Conventional symmetric polynomials are ubiquitous in many areas of mathematics
and mathematical physics. Specially their relation to the representation theory
and theory of invariants makes them to appear in many physical applications (to name a few: Boson-Fermion correspondence,integrable systems, super/string theory, Chern-Simons theory, link invariants and 3-manifolds invariants versus moduli spaces of Riemann surfaces). Here we want to adapt conventional theory to the case of nilpotent commuting $\eta$-variables, having in view the applications to multiqubit systems.

We shall call a $F(\vec{\eta})$  the symmetric $\eta$-function (symmetric $\eta$-polynomial) if
\begin{equation}
F(\eta_{\sigma(1)}, \eta_{\sigma(2)},\dots ,\eta_{\sigma(n)})= F(\eta_1, \eta_2,\dots, \eta_n), \quad \sigma\in S_n
\end{equation}
Then we can consider the elementary symmetric $\eta$-polynomials $e_k=e_k(\vec{\eta})$
\begin{eqnarray}
e_0&=&1\\
e_1&=&\sum_{i=1}^n \eta_i=\mathcal{T}_n\\
e_2&=&\sum_{i<j}^n \eta_i\eta_j\\
&&\dots\\
e_n&=&\eta_1\eta_2\dots\eta_n=\mathcal{E}_n
\end{eqnarray}
From above definition and formulas for powers of $F(\vec{\eta})$ we have
relations
\begin{equation}
e_1^k=k!\,e_k,\quad \mbox{or}\quad e_1e_{k-1}=k\,e_4
\end{equation}
moreover $\Theta({e_i})=e_{n-i}$. Hence, for the even $n$ there exists selfdual
polynomial
$e_{\frac{n}{2}}$. The set of symmetric $\eta$-polynomials is naturally
$\mathbb{Z}_n$ -graded, by the degree of polynomial. An arbitrary symmetric
$\eta$-polynomial $s_k$ has expansion
$s_k(\vec{\eta})=\sum_{i, I_i}s_{I_i}\eta_{I_i}$.
An $s_k$ is homogenous of degree $k$ if
$s_k(\lambda \vec{\eta})=\lambda^k s_k(\vec{\eta})$, where
$\lambda\in\mathbb{R}$.
As in the conventional case we can introduce the analog of  Euler operator counting the degree of the homogenous polynomial. Namely,
\begin{equation}
\vec{\eta}\cdot\nabla=\sum_{i}\eta_i\partial_i,
\end{equation}
where $\partial_i$ denotes derivative with respect to the $\eta_i$ variable (\cite{amf-ncm}) (cf. also the next section). In the context of nilpotent quantum
mechanics we shall call this mapping - the qubit number operator.\\
Let us formulate the fundamental theorem of symmetric $\eta$-polynomials:
the $F(\vec{\eta})$ is symmetric polynomial iff it can expanded in elementary symmetric polynomials
\begin{equation}
F(\vec{\eta})=F(e_1, e_2,\dots, e_n)=\sum_{k=0}^n F_{|k|}e_k=
\sum_{k=0}^n \frac{1}{k!}F_{|k|}e_1^k
\end{equation}
In analogy to conventional theory we can define the complete symmetric $\eta$-polynomials
\begin{equation}
h_k(\vec{\eta})=\sum_{d_1+d_2+\dots +d_n=k}\eta_1^{d_1}\eta_1^{d_2}\dots \eta_n^{d_n},
\end{equation}
where $d_i=0,1$. Using real parameter $t$ they are generated by the function
\begin{equation}
H_n(t)=\sum_{k\geq 0}h_k(\vec{\eta})t^k=\sum_{d_1,d_2,\dots ,d_n}\eta_1^{d_1}\eta_1^{d_2}\dots \eta_n^{d_n}t^{d_1+d_2+\dots +d_n}=
\frac{1}{\prod_{i=n}^n(1-t\eta_i)}
\end{equation}
Analogously, generating function of elementary symmetric $\eta$-polynomial $e_k(\vec{\eta})$
is given as
\begin{equation}
E_n(t)=\sum_{k=0}^n e_k(\vec{\eta})t^k=\sum_{k=0}^n \frac{1}{k!}e_1^k(\vec{\eta})t^k=\exp{t(\eta_1+\eta_2+\dots \eta_n)}=\prod_{k=1}^n e^{t\eta_k}
\end{equation}
Therefore
\begin{equation}
E_n(-t)H_n(t)=1.
\end{equation}
Hence, for $m\geq 1$
\begin{equation}
\sum_{k=0}^n (-1)^ie_k h_{m-i}=0
\end{equation}
As in conventional case one can write above relations as vanishing determinants.\\
Examples:
\begin{itemize}
\item m=1. $e_0 h_1-e_1h_0=0$
\item m=2.
$$
h_2=e_1^2-e_2=\left|
\begin{array}{cc}
e_1 & e_2\\
1 & e_1
\end{array}
\right|
$$
and hence $2e_2=e_1^2$
\end{itemize}
For arbitrary $k$, the identities for the $e_k$ now read as
\begin{equation}
\left|
\begin{array}{ccccc}
e_1 & e_2&\dots &e_{k-1}&e_k \\
1& e_1&\dots &\dots& e_{k-1}\\
0&1&\dots &\dots&e_{k-2}\\
0 & \dots&\dots &1&e_1
\end{array}
\right|
\end{equation}
Finally let us observe, that the notion of antisymmetric $\eta$-polynomial
$a_n(\eta_{\sigma(1)}, \eta_{\sigma(2),\dots},\eta_{\sigma(n)})=(-1)^{|\sigma|}
a_n(\eta_1, \eta_2,\dots, \eta_n)$
is almost trivial. Namely, the Vandermonde determinant
$\Delta(\eta_1, \eta_2,\dots, \eta_n)=\Pi_{1\leq i\leq j\leq n}(\eta_i-\eta_j)$
for nilpotent commuting variables can be different form zero only for $n=2$.
%
\subsection{$\eta$-calculus}
\begin{itemize}
\item[] \underline{\bf $\eta$-derivative}.
The $\eta$-derivative in form we need in the present approach was presented in Ref.\cite{amf-ncm}. To fix the notation let us recall its main properties. Namely, let
\begin{equation}
\partial_j=\frac{\partial}{\partial \eta^j}
\end{equation}
then
\begin{equation}
\partial_i\eta^j=\delta_i^j, \quad \partial_i 1=0,\quad \partial_i\partial_j=
\partial_j\partial_i
\end{equation}
Instead of the Leibniz rule  for $F(\vec{\eta}),\, G(\vec{\eta})\in F[\vec{\eta}]$ we
 have the following relation (para-Leibniz roule \cite{fik})
\begin{equation}
\partial_i(F\cdot G)= \partial_i F\cdot G + G \cdot \partial_i F -
2\eta_i\partial_i F\partial_i G
\end{equation}
For further details cf. Ref. \cite{amf-ncm,amf-nf}.
\item[] \underline{\bf $\eta$-integration}.
The $\eta$-integral is defined by the following contractions on basics variables
\begin{equation}
\int\eta^i d\eta_j=\delta^i_j, \quad \int d\eta_i=0
\end{equation}
and by linearity is extended to the $F_0[\vec{\eta}]$.
Despite the fact that definition is exactly the same as for the Berezin integral, the properties of the multiplication in the algebra $\mathcal{N}$
make that in this formalism the integration by part formula has different  form. Namely,
\begin{equation}
\left(\int Fd\eta\right)\left(\int G\,d\eta\right)=\frac{1}{2}\left(
\int(\partial F)\cdot G
d\eta +  \int F\cdot(\partial G)d\eta \right).
\end{equation}
Detailed properties of this integral are given in Ref.\cite{amf-ncm}.
Below we recall Gaussian integral (firstly considered by Palumbo \cite{palu-1})
which we shall use to define scalar product in $F_0[\vec{\eta}]$. Then we
introduce the $\eta$-Fourier transform. Finally we show the form of generating $\eta$-functions for Stirling numbers of second kind and the $\eta$-integral
form of the triangular recurrence relation for $S(n,k)$. Moreover we give
$\eta$-integral expression for both types of Bell numbers. Because in the definition of the $\eta$-Hermite polynomials we make use of the notion of $\eta$-derivative, therefore we introduce it in this paragraph.
\item[]\underline{{\bf Gaussian $\eta$-integral}.}
Let $B$ be a $n\times n$ symmetric matrix then
\begin{equation}
\int e^{\eta B\eta'}d\vec{\eta}d\vec{\eta'}=per(B),
\end{equation}
where $per(B)$ is the permanent of the square matrix $B$ (cf. \ref{perm}) and  $d\vec{\eta}=d\vec{\eta}_n=d\eta_1d\eta_2\dots d\eta_n$
\item[] {\underline{{\bf$\eta$-Fourier transformation}.}}
Let us define $\eta$-exponent as follows
\begin{equation}
e^{<\vec{\xi},\, \vec{\eta>}}=\sum_k \frac{1}{k!}<\vec{\xi},\, \vec{\eta}>^k, \;\; <\xi,\,
\eta>=\xi^i\eta_i,\;\; (\xi^i)^2=(\eta_i)^2=0.
\end{equation}
A Fourier $\eta$-transform of a function $g(\vec{\eta})\in F[\vec{\eta}]$ we shall
call the function $\hat{g}(\vec{\xi})\in F[\vec{\xi}]$ such that
\begin{equation}
(\mathcal{F}g)(\vec{\xi})=\hat{g}(\vec{\xi})=\int e^{<\vec{\xi},\, \vec{\eta}>}
g(\vec{\eta})d\vec{\eta}.
\end{equation}
Then the inverse Fourier $\eta$-transform is given in the following form
\begin{equation}
\mathcal{F}^{-1}f(\vec{\eta})=\int e^{-<\vec{\xi},\,
\vec{\eta}>}f(\vec{\xi})d\vec{\xi}
\end{equation}
Such defined Fourier $\eta$-transform has properties similar to the Fourier
transform considered in superanalysis. For example
\begin{equation}
\partial_{\xi^i }(\mathcal{F}g)(\vec{\xi})=\mathcal{F}(\vec{\eta}_i g)(\vec{\xi})
\end{equation}
\begin{equation}
\mathcal{F}(f(\eta-\eta_a))(\xi)=e^{<\vec{\xi},\,
\vec{\eta}_a>}(\mathcal{F}f)(\vec{\xi})
\end{equation}
\begin{equation}
(\mathcal{F}f)(\xi+\xi_a))=\mathcal{F}(e^{<\vec{\xi}_a,\, \vec{\eta}>}f)
(\vec{\xi})
\end{equation}
There is also interesting property of Fourier $\eta$-transform obtained when we
want to transform the $\eta$-coordinates. Let $A$ be invertible $n\times n$ matrix
with entries from $\mathcal{N}$ then realizing permutation and scaling transformation
\begin{equation}
\mathcal{F}(A\vec{\eta})(\vec{\xi})=\mathcal{F}f(A^{-1T}\vec{\xi})\cdot per(A),
\end{equation}
 In opposite to the case of the Grassmannian Fourier transform, we do
not obtain here the determinant but permanent. This is characteristic property
of $\eta$-calculus. In particular, for $n=2, 4$ we have that
\begin{equation}
\mathcal{F}(A\vec{\eta})(\vec{\xi})=\mathcal{F}f(A^{-1T}\vec{\xi})\cdot
H\!f^2(A),
\end{equation}
(cf.  \ref{perm}.).
\item[]{\underline{{\bf$\eta$-integral form of the Stirling numbers}.}}
Using directly the expansion of $\mathcal{E}_n$ we can represent the Stirling numbers of the second kind by the following $\eta$-integral
\begin{equation}
S(n,k)=\frac{1}{k!}\int(\mathcal{E}_n)^k d\vec{\eta}_n
\end{equation}
It is easy to see that triangular recurrence relation for $S(n,k)$
\begin{equation}
S(n,k)=k\, S(n-1,k)+S(n-1, k-1)
\end{equation}
can be expressed by means the following compact $\eta$-integral relation
\begin{equation}
\int(\mathcal{E}_n)^k d\vec{\eta}_n=k\,\int(\mathcal{E}_{n-1})^{k-1}e^{\mathcal{T}_{n-1}} d\vec{\eta}_{n-1}
\end{equation}
\item[]{\underline{{\bf$\eta$-integral form of the Bell numbers}.}}
Such form of the normal Bell numbers is direct consequence of the expansion (\ref{Bell2}) and was firstly given in \cite{staples}
\begin{equation}
B_n=\int e^{\mathcal{E}_n}d\vec{\eta}_{n}
\end{equation}
Here we obtain new formula for ordered Bell numbers. Using (\ref{Bell}) and relation
\begin{equation}
2-e^{\mathcal{T}_n}=1-\mathcal{E}_n
\end{equation}
we get
\begin{equation}
C_n=\int (2-e^{\mathcal{T}_n})^{-1}d\vec{\eta}_{n}
\end{equation}
The last formula resembles relevant conventional representation of the Bell numbers.
\item[]{\underline{{\bf$\eta$-Hermite polynomials}.}} Let
$<\!\!\!<\vec{\eta},\,\vec{\eta}>\!\!\!> =\frac{1}{2}e_1^2(\vec{\eta})$. By the $I_k$-th $\eta$-Hermite polynomial we shall understand function of the form
\begin{equation}
H_{I_k}=(\eta_1,\eta_2,\dots, \eta_n)=(-1)^k e^{<\!\!\!<\vec{\eta},\,\vec{\eta}>\!\!\!>}\partial_{I_k}e^{-<\!\!\!<\vec{\eta},\,\vec{\eta}>\!\!\!>},
\end{equation}
where the multi-index $I_k$ is fixed. Now, the $k$-th degree $\eta$-Hermite polynomial is
\begin{equation}
H_{|k|}=\sum_{I_k} H_{I_k},
\end{equation}
where the length $k$ of the multi-indices $I_k$ is fixed. $H_{|k|}$ are symmetric polynomials e.g. for $n=2$: $H_{|0|}=1$, $H_{|1|}=e_1$, $H_{|2|}=-e_0-e_1$; for $n=3$: $H_{|0|}=1$, $H_{|1|}=2e_1+6e_3$, $H_{|2|}=-3e_0-3e_2$, $H_{|3|}=0$.
\end{itemize}
%
%
\section{Qubit systems in $\mathcal{N}$-Hilbert space}\label{n-hilb}
To describe qubit and many-qubit states let us use the structure of bimodule
over $\mathcal{N}$ algebra. Let $\mathcal{H}$ be such $\mathcal{N}$-module. We equip it with the $\mathcal{N}$-scalar product i.e. $\mathcal{N}$-sesquilinear mapping
\begin{equation}
<.\,,.>: \mathcal{H}\times \mathcal{H}\mapsto \mathcal{N}
\end{equation}
such that for $F,G\in \mathcal{H}$
\begin{eqnarray}
<\nu F,G>&=& <F,\nu G>, \quad \nu\in \mathcal{N} \\
<F,G>&=&0 \quad\forall G\in \mathcal{H} \Rightarrow F=0\\
b(<F,G>)^*&=&b(<G,F>)\\
b(<F,F>)&\geq& 0, \quad\forall F \in \mathcal{H}\\
\end{eqnarray}

We shall call such module with $\mathcal{N}$-scalar product the $\mathcal{N}$-Hilbert space. Vectors $F$ with $b(<F,F>)\neq 0$ we shall call physical.
Particular realization of such $\mathcal{N}$-module is given by the function space $F[\vec{\eta}_n]$. In this module we introduce $\mathcal{N}$-valued weakly non-degenerated scalar product in the following form
\begin{equation}
<F,~G>_{\mathcal{N}}=\int F^*(\vec{\eta})G(\vec{\eta})e^{<\vec{\eta}^*,\vec{\eta}>}~d\vec{\eta}^*~d\vec{\eta},
=\int F^*(\vec{\eta})G(\vec{\eta})d\mu(\vec{\eta}^*,~\vec{\eta})
\end{equation}
where
\begin{equation}
F^*(\vec{\eta})=\sum_{k=0}^n\sum_{I_k}F^*_{I_k}{\eta^{I_k}}^*
\end{equation}
and $\star$ denotes complex conjugation. For $\mathcal{F}_0[\vec{\eta}_n]$ obviously $F_{I_k}$ are complex numbers. The first order nilpotents $\eta^{i*}$ are algebraically independent from $\eta^{i}$. In components we have
 \begin{equation}\label{scalar}
<F,~G>_{\mathcal{N}}=\sum_{k=0}\sum_{I_k}F^*_{I_k}G_{I_k}
\end{equation}
To see how 1-qubit algebra is realized in this formalism let us take
$\mathcal{F}[\eta]$ (set of $\eta$-functions of one variable).
In particular, $\eta$-scalar product of $F(\eta)$ and $G(\eta)$
functions takes simple form
\begin{equation}
<F,~G>_{\mathcal{N}}=F^*_0G_0+F^*_1G_1
\end{equation}
and in this space realization of the qubit algebra is given in the
following form
\begin{equation}
d^+=\eta~\cdot\;, \quad\quad d=\partial/\partial\eta
\end{equation}
i.e. operators $d^+$ and $d$ are conjugated with respect to the scalar
product given by (\ref{scalar}) moreover
\begin{equation}
\sigma_3=1-2\eta\partial_{\eta}
\end{equation}
is self-conjugated and $[d,\,d^+]_-=\sigma_3$. Hence, we obtain natural realization of the canonical qubit relations Eq.(\ref{cqr1}-\ref{cqr4}), with $N=\eta\partial_{\eta}$ and $[N, d]_-=d$, $[N, d^+]_-=-d^+$.

Despite the scalar product we can define in the $n$-qubit space another
weakly non-degenerate form,  symmetry of which depends on the parity of $n$.
Namely,
the gradation mapping $J$ allows to consider the natural orthogonal projections
\begin{equation}
\pi_{\pm}F(\vec{\eta})=\frac{1}{2}(F(\vec{\eta})\pm
J(F(\vec{\eta})))
\end{equation}
on the even $F_+$ and odd $F_-$ part of the function $F$. Using the
mapping $J$ we can define the following linear weakly
non-degenerate form
\begin{equation}
\omega_n(F,\, G)=\int J(F(\vec{\eta}))\cdot G(\vec{\eta})d\eta_1\dots d\eta_n=\sum_{k, I_k} (-1)^k F_{I_k}G_{I_{n-k}},\,\,\mbox{where}\,\,\, I_k\cup I_{n-k}=I_n
\end{equation}
The $\omega_n$, $n$ is symmetric or antisymmetric, depending on the parity of $n$
\begin{equation}
\omega_n(F,\, G)=(-1)^n \omega_n(G,\, F)
\end{equation}
Let us note here that it is the $\eta$-version of the form obtained in the
tensor product of $\mathbb{C}^2\otimes\dots\otimes\mathbb{C}^2$ from
the antisymmetric form $\epsilon$ in the $\mathbb{C}^2$.
Explicitly, we have
\begin{itemize}
\item[]\underline{\bf n=1:}
\begin{equation}
\omega_1(F,\, G)=F_0G_1-F_1G_0
\end{equation}
\item[]\underline{\bf n=2:}
\begin{equation}
\omega_2(F,\, G)=F_0G_{12}+F_{12}G_0-F_1G_2-F_2G_1
\end{equation}
In particular for basis $\{1,\eta_1,\,\eta_2,\,\eta_1\eta_2\}$ we get
\begin{equation}
\omega_2(1,\eta_1\eta_2)=1,\quad \omega_2(\eta_1, \eta_2)=1
\end{equation}
and in this case we have exactly the form considered by Wallach \cite{wallach-lect}
\item[]\underline{\bf n=3:}
\begin{equation}
\omega_3(F,\, G)=F_0G_{123}+F_{123}G_0+F_{23}G_1-F_1G_{23}+F_{13}G_2-F_2G_{13}+F_{12}G_3
-F_3G_{12}
\end{equation}
Counterpart of the above symplectic form was discussed by Meyer and
Wallach \cite{mey-wall-comp}, where identification of components
$(x_i, y_j)$ $i,j=1,2,3,4$ used there and components of the
$\eta$-function is the following: $x_1=F_0$, $x_2=F_{23}$,
$x_3=F_{13}$, $x_4=F_{12}$, $y_1=F_{123}$, $y_2=F_1$, $y_3=F_2$,
$y_4=F_3$.
\end{itemize}
In the space of $\eta$-functions we can introduce natural family of
projectors related to the decomposition of the $F(\vec{\eta})$ into the
part depending on the fixed $\eta_k$ and independent of it,
$F(\vec{\eta})=F(\eta_1,\eta_2,\dots,\hat\eta_k,\dots,\eta_n)+\eta_k
\tilde{F}(\eta_1,\eta_2,\dots,\hat\eta_k,\dots,\eta_n)$, where hat
 indicates skipped variable. Natural realization of it is given by
$\eta$-derivative operator, due to the identity
\begin{equation}
\partial_k\eta_k+\eta_k\partial_k=1
\end{equation}
we can introduce projectors
\begin{eqnarray}
\pi_{k|0}&=&\partial_k\eta_k\cdot\\
\pi_{k|1}&=&\eta_k\partial_k\cdot
\end{eqnarray}
For a fixed $k$ they are orthogonal and for different indices $k$ they
commute
\begin{eqnarray}
\pi_{k|i}\pi_{k|j}&=&\delta_{ij}\pi_{k|j}, \quad \pi_{k|0}\oplus\pi_{k|1}=id\\
\pi_{k|i}\pi_{l|j}&=&\pi_{l|j}\pi_{k|i}, \quad k\neq l
\end{eqnarray}
For example the decomposition of the $F(\eta_1,\eta_2)$ has the
following form
\begin{eqnarray}\label{fact1}
 F(\eta_1,\,\eta_2)&=&F_0+F_2\eta_2+\eta_1(F_1+F_{12}\eta_2)\equiv F(\eta_2)+\eta_1 \tilde{F}(\eta_2)\\\nonumber
 &=&\pi_{1|0}F(\eta_1,\,\eta_2)+\pi_{1|1}F(\eta_1,\,\eta_2)
 \end{eqnarray}
 or
\begin{eqnarray}\label{fact2}
 F(\eta_1,\,\eta_2)&=&F_0+F_1\eta_1+\eta_2(F_2+F_{12}\eta_1)\equiv F(\eta_1)+\eta_2 \tilde{F}(\eta_1)\\\nonumber
 &=&\pi_{2|0}F(\eta_1,\,\eta_2)+\pi_{2|1}F(\eta_1,\,\eta_2)
 \end{eqnarray}
Moreover, we get the full decomposition of the $F(\eta_1,\,\eta_2)$  using composition of projectors
\begin{eqnarray}
F_0&=&\pi_{1|1}\pi_{2|0} F(\eta_1,\,\eta_2),\\
F_1\eta_1&=&\pi_{1|1}\pi_{2|0} F(\eta_1,\,\eta_2), \\
F_{12}\eta_1\eta_2&=&\pi_{1|1}\pi_{2|1} F(\eta_1,\,\eta_2)
\end{eqnarray}
In the naturay way this formula generalizes to the $n$ variables.

For $n=1$ we can realize the antisymmetric form $\omega_1$ in alternative way by introducing the wedge product of $\eta$-functions as
\begin{equation}
F(\eta)\wedge G(\eta)\equiv (F_0G_1-F_1G_0)\eta,
\end{equation}
then the anti-symmetric form obtained in this way we shall denote $\mathcal{D}_1$
\begin{equation}
\mathcal{D}_1(F,~G)\equiv \int F(\eta)\wedge G(\eta)d\eta
\end{equation}
For one $\eta$ variable $\mathcal{D}_1(F,~G)=\omega_1$. It is possible
to  generalize the $\mathcal{D}_1$ to the $d=2$ case, in the following
sense. The space of $\eta$-functions $F(\eta_1,\,\eta_2)$ is
$2^2$-dimensional, so admits antisymmetric form. Because, a general
$d=2$  $\eta$-function can be decomposed using one of the factorizations (\ref{fact1}) or (\ref{fact2}) then we can define entity $\mathcal{C}_2$,
which is counterpart of the concurrence of the 2-qubit states, as
\begin{equation}
\mathcal{C}_2\equiv \mathcal{D}_1(F(\eta_1),\tilde{F}(\eta_1))=
\mathcal{D}_1({F}(\eta_2),\tilde{F}(\eta_2))=F_0F_{12}-F_1F_2
\end{equation}
The anti-symmetric form $\mathcal{D}_2$ we define in as follows
\begin{eqnarray}
\mathcal{D}_2&\equiv& \mathcal{D}_1(F(\eta_1),\,\tilde{G}(\eta_1))-
\mathcal{D}_1(G(\eta_1),\,\tilde{F}(\eta_1))\\\nonumber
&=&\mathcal{D}_1(F(\eta_2),\,\tilde{G}(\eta_2))-
\mathcal{D}_1(G(\eta_2),\,\tilde{F}(\eta_2))
\end{eqnarray}
%
%
\subsection{Natural bases: monomial, trigonometric}
Having in mind further application to the $n=2,3,4$  qubit pure states  entanglement let us discuss particular bases in the $\mathcal{F}[\vec{\eta}_n]$.
\subsubsection{monomial basis}
The simplest basis in the space of $\eta$-functions is the monomial one consisting of $\{\eta_{I_k}\}_{k=0}^{n}$. It correspond to the so called computational basis in conventional notation, widely used in the literature. The binary
notation used there for describing the elements of the tensor product of
$\mathbb{C}^2$ is simply related to the $\eta$-notation. Namely,
translation of the multi-index for system of $n$ qubits with the "binary" entries $0$, $1$ to the multi-index $I_k$ used in the $\eta$-function expansion is obtained by putting ordinal numbers equal  the position of $1$'s appearing in the binary multi-index e.g.
$(0,0,0,0)\mapsto 0$, $(1,0,0,0)\mapsto 1$, $(0,1,0,0)\mapsto 2$, \dots, $(0,\,1,\,0,\,1)\mapsto (2,4)$, \dots, $(1,1,1,1)\mapsto (1,2,3,4)$ (cf. also   \ref{conversion}).
Using such notation we obtain for example the following expression for the hyperdeterminant of the hyper-matrix $B=(b_{ijk})$, $i,j,k=0,1$
\begin{eqnarray}\nonumber
Det(F)\equiv Det(B)&=&(F_0^2F_{123}^2+F_{3}^2F_{12}^2+F_{2}^2F_{13}^2+F_{1}^2F_{23}^2)+
4(F_{0}F_{23}F_{13}F_{12}\\ \nonumber
&+&F_{1}F_{2}F_{3}F_{123})-2(F_{0}F_{3}F_{12}F_{123}+F_{0}F_{2}F_{13}F_{123}\\ \nonumber
&+&F_{0}F_{1}F_{23}F_{123}
+F_{2}F_{3}F_{13}F_{12}+F_{1}F_{3}F_{23}F_{12}\\
&+&F_{1}F_{2}F_{23}F_{13})
\end{eqnarray}
\subsubsection{trigonometric basis}
More interesting from the point of view of the entanglement, there are trigonometric bases. For the $n=1$ such basis is identical with the monomial one $\{1, \eta\}$, but for higher $n$ we obtain nontrivial functions.
\begin{itemize}
\item[]\underline{\bf n=2:} We take trigonometric function (\ref{trig-2}) with arguments $\eta_1\pm \eta_2$. These functions are orthogonal and when normalized with respect to our $\mathcal{N}$-scalar product they take the following form
\begin{eqnarray}
h_1&=&\frac{1}{\sqrt{2}}\,cos(\eta^1+\eta^2)=\frac{1}{\sqrt{2}}(1-\eta^1\eta^2)=\psi_{GHZ-},\\
h_2&=&\frac{1}{\sqrt{2}}\,cos(\eta^1-\eta^2)=\frac{1}{\sqrt{2}}(1+\eta^1\eta^2)=\psi_{GHZ+}\\
h_3&=&\frac{1}{\sqrt{2}}\,sin(\eta^1+\eta^2)=\frac{1}{\sqrt{2}}(\eta^1+\eta^2)=\psi_{W+},\\ h_4&=&\frac{1}{\sqrt{2}}\,sin(\eta^1-\eta^2)=\frac{1}{\sqrt{2}}(\eta^1-\eta^2)=\psi_{W-},
\end{eqnarray}
 One can recognize here the $\eta$-realization of the "magic" basis for 2-qubit system. The $cos$-states are GHZ-type and $sin$-states are W-type (Bell states up to the particular phases). According to our definition, above states are physical with respect to the introduced before $\eta$-scalar product with "Gaussian measure".
\item[]\underline{\bf n $\geq$ 3:} Above basis can be generalized to higher $n$, it contains $2^n$ elements
\begin{equation}
\left\{
\frac{1}{\sqrt{2^{n-1}}}\cos(\eta_1\pm \eta_2\pm \eta_3\dots\pm \eta_n), \frac{1}{\sqrt{2^{n-1}}}\sin(\eta_1\pm \eta_2\pm \eta_3\dots\pm \eta_n)
\right\}
\end{equation}
Functions entering above basis are normalized.
%
\end{itemize}
%

%
\subsection{$\eta$-kernels}
We can realize the action of an operator in $\mathcal{N}$-Hilbert spaces of $\eta$-functions using "kernels" with respect to the $\eta$-integral.
\begin{equation}
(AF)(\vec{\eta})=\int A(\vec{\eta},\vec{\eta}')F(\vec{\eta}')d\vec{\eta}'
\end{equation}
Taking expansion of the kernel in the following general form
\begin{equation}
A(\vec{\eta},\vec{\eta}')=\sum_{\stackrel{k,I_k}{l, J_l}} A_{I_k|J_l}\eta^{I_k}{\eta'}^{J_l}
\end{equation}
we obtain the following explicit action on components of the function $F$
\begin{equation}
(AF)_{I_k}=\sum_{l, J_l}A_{I_k|J_l}F_{J_{n-l}},
\end{equation}
where $J_{n-l}$ is complementary strictly ordered multi-index i.e. $J_l\cup J_{n-l}=J_n$. In the fundamental $d=1$ case  we have
\begin{equation}
AF(\eta)=(A_{0|0}F_1+A_{0|1}F_0)+(A_{1|0}F_1+A_{1|1}F_0)\eta.
\end{equation}
and one can easily obtain realizations of some operators important in Hilbert $\eta$-space
\begin{eqnarray}
id&=&A_{id}(\eta,\,\eta')=\eta+\eta'=\delta(\eta-\eta')=\delta(\eta'-\eta)\\
\eta\cdot&=&A_{d^+}(\eta,\,\eta')=\eta\eta'\\
\partial_{\eta}\cdot&=&A_{d}(\eta,\,\eta')=1\\
\partial_{\eta}\eta\cdot&=&A_{\pi_0}(\eta,\,\eta')=\eta'\\
\eta\partial_{\eta}\cdot&=&A_{\pi_1}(\eta,\,\eta')=\eta\\
\sigma_3\cdot&=&A_{\sigma_3}(\eta,\,\eta')=\eta'-\eta, \quad \sigma_3=1-2\eta\partial_{\eta}\\
\sigma_1\cdot&=&A_{\sigma_1}(\eta,\,\eta')=e^{\eta'\eta}=cos(\eta-\eta')=ch(\eta+\eta'), \quad \sigma_1=\partial_{\eta}+\eta\\
\epsilon\cdot&=&A_{\epsilon}(\eta,\,\eta')=e^{-\eta'\eta}=cos(\eta+\eta')=ch(\eta-\eta'),\\\nonumber \epsilon&=&-i\sigma_2=\partial_{\eta}-\eta,
\end{eqnarray}
were symbols $\sigma_i$, $i=1,2,3$ are used because above $\eta$ operators play the role of $\sigma$-matrices in $\eta$-realization of the $su(2)$ algebra and $\pi_k$, $k=0,1$ is the projection on the first and respectively on the second term of the $F(\eta)$ expansion. The Hadamard operator (gate), defined by relations $H 1=\frac{1}{\sqrt{2}}(1+\eta)$ and $H\eta=\frac{1}{\sqrt{2}}(1-\eta)$ has the following $\eta$-differential realization
\begin{equation}
H=\frac{1}{\sqrt{2}}(1+\eta+\partial_{\eta}-2\eta\partial_{\eta})
\end{equation}
and its $\eta$-integral kernel is of the form
\begin{equation}
H\cdot=A_{H}(\eta,\,\eta')=\frac{1}{\sqrt{2}}(1-\eta+\eta'+\eta\eta')=
ch(\eta'+\eta)+sh(\eta'-\eta)=cos(\eta'-\eta)+sin(\eta'-\eta)
\end{equation}

%
%
\section{Nilpotent quantum mechanics and $\eta$-Schr\"{o}\-din\-ger equation}\label{sch}
As it is well known from supersymmetric theories, we can realize classically
fermions using anticommuting variables and formalism of supermechanics or pseudomechanics. There we have prequantum  description of e.g. spin systems, using anticommuting $\theta$-variables and using explicitly supersymmetry or not. The superphase space language, graded Hamiltonians, graded Poisson brackets, graded Heisenberg group, graded special functions etc. appear very useful in classification and analysis of properties of such systems. Then applying cannonical quantization we obtain quantum description of such systems. With this respect there are two approaches. One procedure yields after quantization the conventional Hilbert space formalism and conventional quantum mechanics. Here symbolically the Grassmann algebra of $\theta$-variables is traded for Clifford algebra of $\sigma$-matrices (or $\gamma$-matrices in relativistic case). It is so called Casalbuoni/Berezin=Marinov quantization (\cite{casal, berez-mari}). But there is also another version of quantization procedure, which preserves the $\theta$-variables. This one is specially efective when we ask about
representations of supersymmetry, look for multiplets of states and want to use the Feynman path integral for fermions. It can be symbolically named the super-Schr\"{o}\-din\-ger quantization, because we use there superwavefunctions $\psi(x,\theta)$ and generalized Schr\"{o}\-din\-ger equation, involving superderivatives.

In description of qubits we want to develop analogous approach to the super-Schr\"{o}\-din\-ger one. Some elements of it are already known in literature. On the one hand, in the papers by Mandiliara at al.  \cite{mandi,mandi2} there was already used equation
which can be named $\eta$-Schr\"{o}\-din\-ger equation in the logarithmic form. It was written for a restricted set of
functions of nilpotent commuting variables - nilpotentials, and used to address, via system control methods, some questions of the entanglement. On the other hand there exists classical theory based on nilpotent commuting variables introduced in \cite{amf-ncm} which provides configuration and phase space description of nilpotent systems. It is called nilpotent classical mechanics. Moreover another
essential aspect of such theory i.e. path integral formalism, was discussed some time ago by Palumbo at al. \cite{palu-1, palu-2, palu-3}. Let us note that in all this approaches, except the nilpotent classical mechanics there was neglected the fact, that the derivative with respect to the nilpotent commuting variables do not satisfy the Leibniz rule, what makes the whole construction nontrivial. So, it is natural to consider the nilpotent quantum mechanics as formalism which is related by a "$\eta$-canonical quantization" to the classical nilpotent mechanics. Because known by now $\eta$-Poisson brackets do not satisfy the Jacobi identity, the term
"$\eta$-canonical quantization" leaves some open questions, but the formalism of nilpotent quantum mechanics itself is consistent and effective.  We shall
use here the restricted $\eta$-Schr\"{o}\-din\-ger quantization in the following sense. To quantize classical nilpotent system, we take a classical observable in the normal ordered form i.e. momentum variables are to the right of the coordinate variables and realize position and momentum as operators
\begin{equation}
\eta_k \longrightarrow \hat{\eta}=\eta_k\cdot,\quad\quad p\longrightarrow \hat{p}_k=\frac{\partial}{\partial\eta_k}
\end{equation}
in the $\mathcal{N}$-Hilbert space of $\eta$-functions depending on $\eta_k$, $k=1,2,\dots,n$.
Let $\psi\in\mathcal{F}[x, \vec{\eta}]$
\begin{equation}\label{schro}
i\hbar\frac{d}{dt}\psi(x,\vec{\eta},t)=\hat{H}\psi(x,\vec{\eta},t),
\end{equation}
where $\hat{H}$ is quantized Hamiltonian $H(x,p_x,\eta, p_{\eta}, t)$ of the
system. For the two level systems it is typical to consider explicit time
dependence of the Hamiltonian. For example in the $n=1$ case the Hamilton function is singular in its
nilpotent part in the sense that it contains terms linear in $p_{\eta}$ i.e.
$H=\frac{1}{2m}p_x^2+b(t)p_{\eta}+c(t)\eta p_{\eta}+ V(x,\eta, t)$. After quantization we can write this Hamiltonian in the convenient form
$\hat{H}=\frac{1}{2m}\hat{p}_x^2+V(x)+\vec{B}(t)\cdot\vec{\sigma}$, where nilpotent part can be written as
\begin{equation}
\hat{H}_{nilp}= (B_x(t)+iB_y(t))\eta+(B_x(t)-iB_y(t))\frac{\partial}{\partial \eta} -2B_z(t)\eta\frac{\partial}{\partial \eta}+B_z.
\end{equation}
In the present paper we restrict ourselves and analyze the
properties of the nilpotent part alone, putting aside the question of the
simultaneous $x$ coordinate dependence. We will  assume as well the global
factorization of time dependence of the $\eta$ - wavefunction
$\psi(\vec{\eta},t)$
and study the stationary $\eta$-Schr\"{o}\-din\-ger equation for nilpotent quantum
system.
\begin{equation}
\hat{H}\psi(\vec{\eta})=\lambda \psi(\vec{\eta})
\end{equation}
The structure of eigenstates for multiqubit systems turns out to bo nontrivial,
when one addresses the question of entanglement.

Let us compare considered above the $\eta$-Schr\"{o}\-din\-ger equation with the one
studied by Mandiliara et all \cite{mandi}. In the latter one, the authors restrict to the case when $\eta$ - wave function has invertible values in the algebra $\mathcal{N}$ (we describe situation using formalism developed in the present work) i.e. $\psi(\vec{\eta})=\psi_0+\psi_i\eta_i+ \dots$ and $\psi_0\neq 0$ therefore one can take function $\tilde{\psi}(\vec{\eta})=\frac{1}{\psi_0}\psi(\vec{\eta})$ and there exists its logarithm $f(\vec{\eta})=ln\tilde{\psi}$. Now, because
\begin{equation}
i\frac{d}{dt}f(\vec{\eta})=i\frac{d}{dt}ln\tilde{\psi}(\vec{\eta})=
i\tilde{\psi}^{-1}(\vec{\eta})H\tilde{\psi}(\vec{\eta})
\end{equation}
and $\tilde{\psi}(\vec{\eta})=e^{f(\vec{\eta})}$ we get
\begin{equation}
i\frac{d}{dt}f(\vec{\eta})=e^{-f(\vec{\eta})}He^{f(\vec{\eta})},
\end{equation}
what is the form of the equation employed in \cite{mandi}. But let us note once
again, there many other states that have $\eta$ - wave functions with
non-invertible values (like Werner -like states ) and for them such equation is
not valid, but $\eta$-Schr\"{o}\-din\-ger equation (\ref{schro}) can be used without problem.
%
%
\section{Two component nilpotent systems}\label{2comp}
%

Let us consider well known generic Hamiltonian for the two, two level quantum systems. In terms of $\sigma$ - matrices it is given by
\begin{equation}
\hat{H}_{c_1c_2c_3}=c_1\sigma^x\otimes\sigma^x+c_2\sigma^y\otimes\sigma^y+c_3\sigma^z\otimes\sigma^z,
\end{equation}
where $c_i$, $i=1,2,3$ are numerical parameters.
In the $\eta$-Schr\"{o}\-din\-ger representation it can be written in the following
form
\begin{eqnarray}
\hat{H}&=&(c_1-c_2)(d^+\otimes d^+ +d\otimes d)+(c_1+c_2)(d^+\otimes d+d\otimes d^+)\\\nonumber
&-&2c_3(d^+d\otimes 1+1\otimes d^+d)
+4c_3(d^+d\otimes d^+d+\frac{1}{4}),
\end{eqnarray}
where as before $d^+=\eta\cdot$ and $d=\partial_{\eta}$. There are several special choices of the values of the parameters discussed in the literature:
\begin{enumerate}
\item Ising coupling: $c=c_1\neq 0$, $c_2=c_3=0$
\begin{equation}
\hat{H}_{Is}=c(d^+\otimes d^+ + d\otimes d)+c(d^+\otimes d+d\otimes d^+)
\end{equation}
It is interesting to note, that classical Hamiltonian for this system is composed of $\eta$-harmonic oscillator and additional angular momentum like term i.e. $H=c(p_1p_2+\eta_1\eta_2+\eta_1p_2+\eta_2p_1)$
\item XY coupling: $c=c_1=c_2\neq 0$, $c_3=0$
\begin{equation}
\hat{H}_{XY}=2c(d^+\otimes d+d\otimes d^+)
\end{equation}
Here the $\eta$-harmonic oscillator part is not present
\item Heisenberg (spin system) type: $c=c_1=c_2$, $c_3\neq 0$
\begin{equation}
\hat{H}_{He}=2c(d^+\otimes d + d\otimes d^+)-2c_3(d^+d^+\otimes {1}+d\otimes d^+d)+4c_3(d^+d\otimes d^+d+\frac{1}{4})
\end{equation}
Here there it is a subcase with $c=0$.
\item $Q$-invariant type: $c=c_1=-c_2$
\begin{equation}
\hat{H}_{Q}=2c(d^+\otimes d^+ + d\otimes d)-2c_3(d^+d^+\otimes 1+d\otimes d^+d)+4c_3(d^+d\otimes d^+d+\frac{1}{4})
\end{equation}
\end{enumerate}
Before solving the $\eta$-eigenfunction problem, let us observe that for
the $\hat{H}_{c_1c_2c_3}$ family of Hamiltonians there exist conserved charges.
One is related to the duality transformation $\Theta$. For the two-qubit system the $\Theta$ is implemented by operator
\begin{equation}
\Theta^{(2)}=d\otimes d+ d^+\otimes d + d\otimes d^++d^+\otimes d^+.
\end{equation}
We have that
\begin{equation}
[\Theta^{(2)},\,\hat{H}_{c_1c_2c_3}]_-=0, \quad\quad \forall c_1,\,c_2,\,c_3
\end{equation}
Let us define two charges
\begin{eqnarray}
Q=d\otimes d^+\\
Q^+=d^+\otimes d
\end{eqnarray}
The $Q$ and $Q^+$ are conserved only for the $\hat{H}_{c_1c_2c_3}$ with $c_1+c_2=0$.

Let us now study in the $\eta$-Schr\"{o}\-din\-ger formalism, the structure of the eigenstates of this composite system. The $\hat{H}_{c_1c_2c_3}$ Hamiltonian has explicitly the following realization in $\mathcal{F}_0(\vec{\eta})$
\begin{eqnarray}
\hat{H}_{c_1c_2c_3}&=&((c_1+c_2)\eta_2-2c_3\eta_1)\partial_1
+((c_1+c_2)\eta_1-2c_3\eta_2)\partial_2\\\nonumber
&+&(c_1-c_2)\eta_1\eta_2+
((c_1-c_2)+4c_3\eta_1\eta_2)\partial_1\partial_2+c_3
\end{eqnarray}
For the eigenproblem
\begin{equation}
\hat{H}_{c_1c_2c_3}\psi(\eta_1,\,\eta_2)=\lambda\psi(\eta_1,\,\eta_2)
\end{equation}
the set of solutions is shown in the Table \ref{Eig_n2}.
%
\begin{table}
\caption{Eigensystem for $n=2$ ($\psi_0$, $\psi_1$, $\psi_2$, $\psi_{12}$ are arbitrary).}\label{Eig_n2}
\begin{center}
\begin{tabular}{llllll}
\hline\\[-5mm]
\hline\\[-2mm]
$\lambda_1=c_1-c_2+c_3$ & & $\Psi_1=\frac{1}{\sqrt{2}}(1+\eta_1\eta_2)$ & &
$c_1-c_2\neq 0$\\[1mm]
& &$\tilde{\Psi}_1=\psi_0+\psi_{12}\eta_1\eta_2$ & & $c_1-c_2 = 0$\\
\hline\\[-2mm]
$\lambda_2=-(c_1-c_2)+c_3$ & & $\Psi_2=\frac{1}{\sqrt{2}}(1-\eta_1\eta_2)$ & &
$c_1-c_2\neq 0$\\
& &$\tilde{\Psi}_2=\psi_0+\psi_{12}\eta_1\eta_2$ & & $c_1-c_2 = 0$\\
\hline\\[-2mm]
$\lambda_3=-(c_1+c_2)-c_3$ & & $\Psi_3=\frac{1}{\sqrt{2}}(\eta_1-\eta_2)$ & &
$c_1+c_2\neq 0$\\
& &$\tilde{\Psi}_3=\psi_1\eta_1+\psi_{2}\eta_2$ & & $c_1+c_2 = 0$\\
\hline\\[-2mm]
$\lambda_4=c_1+c_2-c_3$ & & $\Psi_4=\frac{1}{\sqrt{2}}(\eta_1+\eta_2)$ & &
$c_1+c_2\neq 0$\\[1mm]
& &$\tilde{\Psi}_4=\psi_{1}\eta_1+\psi_2\eta_2$ & & $c_1+c_2 = 0$\\[1.5mm]
\hline\\[-5mm]
\hline
\end{tabular}
\end{center}
\end{table}
For $c_1-c_2=0$ there is degeneration with $\lambda_1=\lambda_2$ and analogously $\lambda_3=\lambda_4$ for $c_1+c_2=0$. Because the duality transformation commutes with the Hamiltonian the eigenvectors have to be selfdual or antiselfdual i.e
$\Theta(\psi_1)=\psi_1$, $\Theta(\psi_2)=-\psi_1$, $\Theta(\psi_3)=-\psi_3$ and $\Theta(\psi_4)=\psi_4$. A nontrivial action of duality transformation is in eigenspaces with degenerate $\lambda$.

Besides the considered above qubit-qubit system, we shall discuss the system with supersymmetry which is
composed of qubit and fermion (Q-F) in analogy to the standard case of boson-fermion (B-F) system. In addition we consider below the boson-qubit (B-Q)
and fermion-fermion (F-F) systems for which at least one part of the composed system is nilpotent. In these cases one can add to the conventional SUSY-like
Hamiltonian the F\"orster term, which gives exchange interaction and is characteristic for the two level systems \cite{mah}, but breaks the SUSY.
%
\subsection{boson-fermion}
To fix the notation let us recall the well known SUSY system of
quantum bosonic and fermionic oscillators. Hamiltonian of such a composed
system  has simple form
\begin{equation}
H^{(0)}=\omega ~(1\otimes f^+f +b^+ b\otimes 1)=\omega \left(1\otimes (f^+f-\frac{1}{2})
+(b^+ b+\frac{1}{2})\otimes 1\right)
\end{equation}
It is a special case of more general Jaynnes-Cummings Hamiltonian which
is supersymmetric only for special values of parameters. Namely, for
$\omega=\omega_0$ and $\kappa=0$
\begin{equation}
H_{JC}=\omega ~(b^+ b-\frac{1}{2})\otimes 1+\omega_0 ~1\otimes (f^+f-\frac{1}{2})+
\kappa H^{(F)}\, ,
\end{equation}
where $H^{(F)}$ is the F\"orster Hamiltonian \cite{mah} of the form
\begin{equation}
H^{(F)}=b^+\otimes f + b\otimes f^+
\end{equation}
To have consistent grading the $H^{(F)}$ and $\kappa$ parameter are odd entities.
Let us write explicitly SUSY algebra for the Jaynnes-Cummings system. The
supercharges are defined as
\begin{equation}
Q_{BF}=i\sqrt{\omega}~b\otimes f^+,\quad Q^+_{BF}=-i\sqrt{\omega}~b^+\otimes f,
\end{equation}
and hamiltonian has the form
\begin{equation}
H^{(F)}_{BF}=Q^+_{BF}+Q_{BF}, \quad H_{BF}= H^{(0)}_{BF}+\kappa H^{(F)}_{BF}
\end{equation}
Graded commutation relations now read as
\begin{equation}
[Q^+_{BF},~Q_{BF}]_+= H^{(0)}_{BF},
\end{equation}
\begin{equation}
[H_{BF},~Q^+_{BF}]_-=\kappa H^{(0)}_{BF}, \quad
[H_{BF},~Q_{BF}]_-=\kappa H^{(0)}_{BF}
\end{equation}
%
\subsection{qubit-fermion}\label{q-f}
For this system composed of qubit and fermion we have the odd supercharges as in the conventional boson-fermion case
\begin{equation}
Q_{QF}=i\sqrt{\omega}~d\otimes f^+,\quad Q^+_{QF}=-i\sqrt{\omega}~d^+\otimes f
\end{equation}
Graded commutation relations now read as
\begin{equation}
[Q^+_{QF},~Q_{QF}]_+=\omega \left( d^+d\otimes 1+(1-2N_d)\otimes f^+f\right)=H^{(0)}_{QF}
\end{equation}
\begin{equation}
H^{(F)}_{QF}=Q^+_{QF}+Q_{QF},
\quad\quad H_{QF}= H^{(0)}_{QF}+\kappa H^{(F)}_{QF}
\end{equation}
\begin{equation}
 [H_{QF},~Q_{QF}]_-=\kappa H^{(0)}_{QF},
\quad \quad [H_{QF},~Q^+_{QF}]_-=\kappa H^{(0)}_{QF}
\end{equation}
This system is very interesting because both of its parts are two level systems and we still have nontrivial supersymmetry transformations with graded algebra of charges (for $H^{(F)}_{QF}=0$).\\
The Hamiltonian $H^{(0)}_{QF}$ in Schr\"{o}\-din\-ger representation takes the form
\begin{equation}
\hat{H}^{(0)}_{QF}=\omega(\eta\partial_{\eta}+\theta\partial_{\theta}
-2\eta\theta\partial_{\eta}\partial_{\theta})
\end{equation}
Solving generalized stationary $\eta$-Schr\"{o}\-din\-ger equation
$\hat{H}^{(0)}_{QF}F(\eta, \theta)=\lambda F(\eta, \theta)$
we get that there is eigenspace related to the zero energy, and another one to $\lambda=\omega$. So the vacuum is degenerated and invariant under supersymmetry transformations. The subspace with nonzero energy is degenerated, as it should be in supersymmetric system. The non-unique ground state is peculiar.
So, we have $\phi_0=1$ and $\psi_0=\eta\theta$ even and odd respectively, ground  states
(in the sense of the Grassmannian parity) and $\phi_{\omega}=\eta$, $\psi_{\omega}=\theta$ even and odd excited states.
\begin{eqnarray}
\hat{H}^{(0)}_{QF}\phi_{\omega}&=&\omega\phi_{\omega},\quad
\hat{H}^{(0)}_{QF}\psi_{\omega}=\omega\psi_{\omega},\quad
\hat{H}^{(0)}_{QF}\phi_0=0,\quad\hat{H}^{(0)}_{QF}\psi_0=0;\\
\hat{Q}\phi_{\omega}&=&\psi_{\omega},\quad \hat{Q}\psi_{\omega}=0,\quad \hat{Q}\phi_0=0,\quad\hat{Q}\psi_0=0\\
\hat{Q}^+\phi_{\omega}&=&0,\quad \hat{Q}^+\psi_{\omega}=\phi_{\omega},\quad \hat{Q}^+\phi_0=0,\quad\hat{Q}^+\psi_0=0
\end{eqnarray}
Spectrum of the qubit-fermion supersymmetric system can be depicted in the following diagram\\
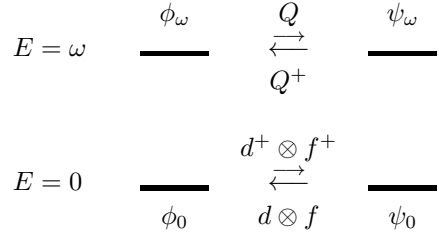
\begin{center}
\begin{tabular}{lccc}
&$\phi_{\omega}$&$Q$&$\psi_{\omega}$\\
$E=\omega$\rule{3mm}{0mm}&\rule{9mm}{0.6mm}&$\stackrel{\longrightarrow}{\longleftarrow}$&\rule{9mm}{0.6mm}\\
&&$Q^+$&\\
&&&\\
&&$d^+\otimes f^+$&\\
$E=0$\rule{3mm}{0mm}&\rule{9mm}{0.6mm}&$\stackrel{\longrightarrow}{\longleftarrow}$&\rule{9mm}{0.6mm}\\
&$\phi_0$&$d\otimes f$&$\psi_0$
\end{tabular}\\[6mm]
\end{center}
In the above example we use convention that $\phi_i$ are even and $\psi_i$ are
odd, $i=0,\omega$.

In terms of the supersymmetric quantum mechanics, we have here
pair of nontrivial zero modes, and therefore the full even and odd spectra are identical.
The analog of Witten index vanishes i.e. $\Delta=n^{(E=0)}_Q-n^{(E=0)}_F=0$.
This might be surprising because, in
conventional supersymmetric quantum mechanics even and odd spectra coincide except the zero energy ground state. However this effect  is not new and is present in SUSY quantum mechanical models with a periodic potential \cite{du-f, du-m} or with local and nonlocal potentials \cite{cho-ho} as well as in the  model of spin $\frac{1}{2}$ particle in a rotating magnetic field and constant scalar potential \cite{tk-r}.
Here in the qubit-fermion system, the effect  is of algebraical origin and comes from the structure of the model, not from the particular properties of the potential.
%
\subsection{qubit-qubit}
For completeness we rewrite in the same notational convention the $Q$- invariant
two qubit system (considered already above). The $Q$-charges this time are even, but still nilpotent
\begin{equation}
Q_{QQ}=i\sqrt{\omega}~d\otimes d^+,\quad Q^+_{QQ}=-i\sqrt{\omega}~d^+\otimes d
\end{equation}
and graded commutation relations now read as
\begin{equation}
[Q^+_{QQ},~Q_{QQ}]_+=\omega \left( d^+d\otimes 1-2N_d) +(1-2N_d)\otimes f^+f\right)=H^{(0)}_{QF}
\end{equation}
\begin{equation}
H^{(F)}_{QQ}=Q^+_{QQ}+Q_{QQ},
\quad\quad H_{QQ}= H^{(0)}_{QQ}+\tilde{\kappa} H^{(F)}_{QQ}
\end{equation}
\begin{equation}
 [H_{QQ},~Q_{QQ}]_-=\tilde{\kappa} H^{(0)}_{QF},
\quad \quad [H_{QQ},~Q^+_{QQ}]_-=\kappa H^{(0)}_{QF}
\end{equation}
%
\subsection{qubit-boson}
\begin{equation}
Q_{QB}=i\sqrt{\omega}~b\otimes d^+,\quad Q^+_{QF}=-i\sqrt{\omega}~b^+\otimes d
\end{equation}
and commutation relations now read as
\begin{equation}
[Q^+_{QB},~Q_{QB}]_-=\omega \left(-b^+b\otimes (1-2N_d)+1\otimes d^+d\right)=
H^{(0)}_{QB}
\end{equation}
\begin{equation}
H^{(F)}_{QB}=Q^+_{QB}+Q_{QB}, \quad H_{QF}= H^{(0)}_{QF}+\lambda H^{(F)}_{QB}
\end{equation}
For this system $H^{(F)}$ and $\lambda$ are even entities
\begin{equation}
 [H_{QB},~Q_{QB}]_-=-\lambda H^{(0)}_{QB},
\quad [H_{QB},~Q^+_{QB}]_-=\lambda H^{(0)}_{QB}
\end{equation}
%
%
\section{Separability and entanglement in $\mathcal{N}$-Hilbert space representation}\label{sep}
The commuting nilpotent variables were already used in the description
of the entanglement in papers of Mandiliara et all. \cite{mandi, mandi2}.
Using advanced in the above sections notions of:  $\eta$-functions, $\eta$-differential calculus and
the $\eta$-formalism of $\mathcal{N}$-Hilbert spaces  we give a general characterization of pure state entanglement in terms of such a variables, and we find states considered in the Ref. \cite{mandi}  related to tanglemeters  as a special case.

As we have already seen, natural bases in tensor
products of composite system are described by elementary functions, which encode
automatically the "combinatorial" content of such non-simple (non-decomposable)
tensors. In the approach presented in \cite{mandi} there are discussed
tanglemeters which are well defined for $\eta$-functions with nonvanishing body.
In addition we shall discuss the case of the $\eta$-function states with trivial
body as well. In such a case one can apply the entanglement monotones introduced
by Meyer and Wallach \cite{mey-wall-comp, mey-wall-jmp} and their generalizations \cite{emary}.

As it is known \cite{ami-faz-ost} one can also consider entanglement of fermions, bosons. In the case of fermions there is no Schmidt decomposition, but an analog of Schmidt rank (Slater rank) can be defined to classify entanglement in bipartite fermionic systems \cite{ec-sc-br-le}.
\subsection{Factorization of $\eta$-functions and pure state entanglement}
In this section we will show, how questions concerning the entanglement can be
answered using the criteria of factorization of $\eta$-functions. We will take
inspiration from the factorization theory of functions of real variable, but of
course proofs of the present theorems come from different reasoning then in
conventional case. To begin consider the simple fact of linear independence of
functions of one $\eta$-variable. This can be interpreted as special case of
factorization $aF(\eta)+bG(\eta)=0\Rightarrow F(\eta)=-\frac{b}{a}G(\eta), \: a\neq 0$. Functions $F(\eta)$ and $G(\eta)$ are linearly dependent iff the Wronskian of the following matrix
\begin{equation}
W=\left(
\begin{array}{cc}
F(\eta)&G(\eta)\\
\partial F(\eta)& \partial G(\eta)
\end{array}
\right),
\end{equation}
$w=F_0G_1-F_1G_0$
vanishes. In particular functions with nonvanishing body, i.e. $F_0\neq 0$ can be written as $F(\eta)=F_0e^{\frac{F_1}{F_0}\eta}$.

Let us define the entity $\mathcal{H}$ which will be important in expressing Wronskians for n-qubit systems. Namely,
\begin{equation}\label{inv H}
\mathcal{H}=\sum_{k=0}^{[\frac{n}{2}]}\sum_{I_k}(-1)^k
\left(\partial_{I_k}F(\vec{\eta})
\partial_{I_{n-k}}F(\vec{\eta})\right)|_{\vec{\eta}=0}
\end{equation}
where $\partial_{I_0}F(\vec{\eta})=\partial_{\emptyset}F(\vec{\eta})=F(\vec{\eta})$.
In particular
\begin{itemize}
\item for $n=2$:
\begin{equation}\label{h2}
\quad\mathcal{H}=F_0F_{12}-F_1F_2,
\end{equation}
\item for $n=3$:
    \begin{equation}\label{h3}
    \quad\mathcal{H}=F_0F_{123}-F_1F_{23}-F_2F_{13}-F_3F_{12},
    \end{equation}
\item for $n=4$:
we get the well known Cayley determinant
\end{itemize}
\begin{equation}\label{h4}
\mathcal{H}=F_0F_{1234}-F_1F_{234}-F_2F_{134}-F_3F_{124}-F_4F_{123}+F_{12}F_{34}+
F_{13}F_{24}+F_{14}F_{23}
\end{equation}
We do not distinguish in the notation the $\mathcal{H}$ for variuos values of the $n$, but it will always be clear from the context which one is under consideration.
%
\subsection{Factorization of $F(\eta_1, \eta_2)$}
As it is known \cite{cyp, ras, nauman, simsa} a sufficiently smooth function of the real variables $f(x,y)$ of the form
$f(x,y)=h(x)g(y)$ has to satisfy the d'Alembert condition i.e.
\begin{equation}\label{dal}
\frac{\partial^2 ln f}{\partial x\partial y}=0.
\end{equation}
This equation can be written in the form
\begin{equation}\label{ww}
\left|
\begin{array}{cc}
f&\frac{\partial f}{\partial x}\\
\frac{\partial f}{\partial y}&\frac{\partial^2 f}{\partial x\partial y}
\end{array}
\right|=0
\end{equation}
But the set of solutions of (\ref{ww}) is larger then factorizable functions
\cite{ras, pra-ras} and there are various generalization of above equation for functions of several variables \cite{nauman, simsa}.

For the $\eta$-functions analog of the  relation (\ref{ww}) is also more general  then that of the d'Alembert equation (\ref{dal}) moreover it is also necessary and sufficient condition for factorability of function. In case when $F(\eta_1, \eta_2)$ takes only invertible values, there exists $f((\eta_1, \eta_2))=lnF(\eta_1, \eta_2)$ and above conditions are equivalent.

Let $w_{12}$ denotes the following $\eta$-Wronskian with respect to $\eta_1$ and $\eta_2$ variables
\begin{equation}\label{w2eta}
w_{12}(F(\eta_1, \eta_2))=det W_{12}=\left|
\begin{array}{cc}
F&\frac{\partial F}{\partial \eta_1}\\
\frac{\partial F}{\partial \eta_2}&\frac{\partial^2 F}{\partial \eta_1\partial \eta_2}
\end{array}
\right|=
\left|
\begin{array}{cc}
F&\partial_1 F\\
\partial_2 F&\partial_{12} F
\end{array}
\right|=F_0F_{12}-F_1F_2
\end{equation}
Let  $F(\eta_1, \eta_2)$ be arbitrary function, then $w_{12}(F)=0$ iff $F(\eta_1, \eta_2)=G(\eta_1)\tilde{G}(\eta_2)$, for some $G$ and $\tilde{G}$. (Proof is
given in   \ref{proof-w2}). Note that the Wronskian for the function of two $\eta$ variables has numerical values (i.e. its soul vanishes and there is no explicit $\eta$ dependence) and $w_{12}=\mathcal{H} (cf. Eq. \ref{inv H})$.

As an example consider the Werner state which is represented by the function
$$
\psi_W(\eta_1, \eta_2)=\frac{1}{\sqrt{2}}(\eta_1+\eta_2),
$$
and we have that  $w_{12}(\psi_W)=-\frac{1}{2}$. For GHZ state
$$
\psi_{GHZ}(\eta_1, \eta_2)=\frac{1}{\sqrt{2}}(1+\eta_1\eta_2)
$$
we get $w_{12}(\psi_{GHZ})=\frac{1}{2}$, but now there exists $ln(\sqrt{2}\psi_{GHZ})=\eta_1\eta_2$ and d'Alembert equation also shows that $\psi_{GHZ}$ is not factorable. Generalization of the d'Alembert equation to the system of $n$ qubits was used by \cite{mandi, mandi2} as a criterion of bipartite entanglement, and by means of the $ln F$ there was defined tanglemeter, as an additive measure of entanglement.

Let us note, that from the condition of vanishing Wronskian we can distinguish several types of factorization: $e^{\eta_1}e^{\eta_2}$, $\eta_ie^{\eta_i}$, $\eta_ie^{\eta_j}$, where $i\neq j$; $i,j=1,2$. On the other hand the non-vanishing Wronskian gives two types of $\eta$-functions: GHZ-like (
$e^{\pm\eta_1\eta_2}$, $1\pm\eta_ie^{\pm\eta_j}$, $e^{\pm\eta_1\eta_2}\pm e^{\eta_i}$) and W-like ($1\pm \eta_1\pm\eta_2$, $\pm \eta_1\pm \eta_2$). This can be seen from the
easy to proof fact that for any function $F(\eta_1, \eta_2)$ there exists factor $f(\eta_1)$ (or $f(\eta_2)$) and function $G(\eta_1, \eta_2)$
such that
\begin{equation}\label{phf}
F(\eta_1, \eta_2)=f(\eta_1)G(\eta_1, \eta_2),
\end{equation}
where
$f(\eta_1)=f_0e^{\frac{f_1}{f_0}\eta_1}$. Taking above $\eta$-function phases with respect to both variables and  assuming that $f_0=1$ and $g_0=1$ we get that
\begin{equation}
F(\eta_1, \eta_2)=f(\eta_1)g(\eta_2)G(\eta_1, \eta_2)=e^{f_1\eta_1+g_2\eta_2}
G(\eta_1, \eta_2)
\end{equation}
and the $w_{12}(F)=w_{12}(G)$. Hence, both functions have the same factorization properties. For example functions related in this way to $\psi_{GHZ}$ have the following form $F(\eta_1, \eta_2)=1+f_1\eta_1+g_2\eta_2+(f_1g_2+1)\eta_1\eta_2$ and these related to $\psi_W$ are: $F(\eta_1, \eta_2)=\eta_1+\eta_2+(f_1+g_2)\eta_1\eta_2$.
%
\subsection{Factorization of $F(\eta_1, \eta_2, \eta_3)$}\label{fac-3}
In the case of three variables we have to consider a set of the Wronski matrices, for all distinct pairs of variables. From such a bipartite information
one can determine the
level of non/factorability of $F(\eta_1, \eta_2, \eta_3)$. This time the Wronskians depend explicitly on $\eta$-variables
\begin{eqnarray}
w_{12}(F)(\eta_3)&=&w_{12}(F|_{\eta_3=0})+(\mathcal{H}+2F_3F_{12})\eta_3
\equiv w_{12}(F|_{\eta_3=0})+\tilde{\mathcal{{H}}}_3\eta_3\\
w_{13}(F)(\eta_2)&=&w_{13}(F|_{\eta_2=0})+(\mathcal{H}+2F_2F_{13})\eta_2
\equiv w_{13}(F|_{\eta_2=0})+\tilde{\mathcal{{H}}}_2\eta_2\\
w_{23}(F)(\eta_1)&=&w_{23}(F|_{\eta_1=0})+(\mathcal{H}+2F_1F_{23})\eta_1
\equiv w_{23}(F|_{\eta_1=0})+\tilde{\mathcal{{H}}}_1\eta_1,
\end{eqnarray}
where $\mathcal{H}$ is given by (\ref{h3})
according to Eq.(\ref{inv H}). What is interesting, one can express all terms in the above expansions of $\eta$-function determinants  in terms of the invariants of Wronski matrices. Namely, we find that the $\tilde{\mathcal{H}}_k(F)$ can be written in terms of traces
\begin{equation}\label{tr-w}
\tilde{\mathcal{H}}_k(F)=tr W_{ij}(F|_{\eta_k=0})\cdot tr W_{ij}(\partial_k F)-
tr\left( W_{ij}(F|_{\eta_k=0}) W_{ij}(\partial_k F)\right)
\end{equation}

Before we  state criterion allowing to detect the existence of factorization (separation) of variables let us observe the following\\[4mm]
{\bf \underline{Weak factorization:}}
Let $F(\eta_1, \eta_2, \eta_3)$ be such, that there exist functions $G(\eta_i, \eta_k)$ and $\tilde{G}(\eta_j, \eta_k)$ and
$F(\eta_1, \eta_2, \eta_3)=G(\eta_i, \eta_k)\tilde{G}(\eta_j, \eta_k)$;
$i,j,k=1,2,3$, are all different and fixed, then $w_{ij}(F)(\eta_k)=0$.\\[3mm]
However, inverse is not true, for example for the $\eta$-function $F(\eta_1, \eta_, \eta_3)=\eta_1\eta_2+\eta_1\eta_3+\eta_2\eta_3$, $w_{ij}(F)(\eta_k)=0$, but there is no weak decomposition $F=(\eta_1, \eta_3)\tilde{G}(\eta_2, \eta_3)$.
The non-vanishing Wronskians excludes possibility of factorization.

One can introduce effective criterion selective enough to indicate full factorization of two subsystems $(i-k)(j)$, with true separation of dependence on one variable.\\[4mm]
{\bf \underline{Strong factorization (separability):}}
Let $F=F(\eta_1, \eta_2, \eta_3)$, there exist functions $G(\eta_i, \eta_k)$ and $\tilde{G}(\eta_j)$ such that
$F(\eta_1, \eta_2, \eta_3)=G(\eta_i, \eta_k)\tilde{G}(\eta_j)$ iff the following conditions are satisfied: $w_{ij}(F)(\eta_k)=0$, $w_{kj}(F)(\eta_i)=0$,
$w_{ij}(\partial_kF)=0$, and $w_{kj}(\partial_i F)=0$ ($i,j,k=1,2,3$, are all different and fixed).\\[3mm]
The proofs of above theorem we present in the   \ref{sft}. When there exists the logarithm of the $\eta$-function then, condition for separability of variables takes very simple form and such criterion was already proposed in \cite{mandi}. Namely, for $n=3$, the dependence on $\eta_j$ variable is separable
iff $\partial_i\partial_j ln(F)=0=\partial_k\partial_j ln(F)$, $i,j,k=1,2,3$.
Moreover, as it is easy to see, the vanishing of above derivatives of the $ln F$ is equivalent to the simple condition: $w_{ik}(F)=0=w_{jk}(F)$ e.g.
$\partial_1\partial_3 ln F(\eta_1,\eta_2, \eta_3)=0 \Leftrightarrow (w_{13}(F|_{\eta_2})=0$ and $\tilde{\mathcal{H}}_2=0$).
In this case, the rest of the conditions present in the strong factorization criterion is fulfilled automatically. Such simple characterization of separability in terms of logarithm generalizes to higher $n$ \cite{mandi}. However, the set of functions with vanishing body $b(F)=0$, and hence not having logarithm, is very large and essential. It grows with the value of $n$. All Werner-like states follow into this type.

From the point of view of the description of physical systems only the strong
factorization gives information of the existing true bipartite separation of
the systems, but even existence of such a factorization still gives room for an another portion of non-separability inside a two-qubit subsystem, except the totaly separable function i.e. $F\tilde = G(\eta_1)\tilde{G}(\eta_2)\dots \tilde{\tilde{G}}(\eta_n)$. The decomposition of $F(\vec{\eta})$ into factors is not unique in general.

Let us note that factorization properties of the $\eta$-function $F(\eta_1, \eta_2, \eta_3)$ and its dual $\star F(\eta_1, \eta_2, \eta_3)$ are closely related, because
\begin{equation}
w_{ij}(F|_{\eta_k=0})=w_{ij}(\partial_k (\star F)), \quad\quad
\tilde{\mathcal{H}}_i(F)=\tilde{\mathcal{H}}_i(\star F)
\end{equation}

To illustrate how non/factorization properties of the $F(\eta_1, \eta_2, \eta_3)$ are encoded in values of considered Wronskians let us consider some examples. For separability the normalization of the function is unimportant, but in view of further applications we shall consider normalized $\eta$-functions.\\
\textbf{\underline{Examples:}}
\begin{enumerate}
\item[{\bf E1:}]
The Werner state is represented by the $\eta$-function $\psi_W=\frac{1}{\sqrt{3}}(\eta_1+\eta_2+\eta_3)$, it is nonseparable, but it is non-factorable in the weak sense as well. We have
\begin{eqnarray}
w_{ij}(\psi_W|_{\eta_k=0})&=&-\frac{1}{3}, \quad\quad \tilde{\mathcal{H}}_i=0,
\\
w_{ij}(\partial_k\psi_W)&=&0
\end{eqnarray}
The cluster Werner state is represented by the $\eta$-function dual to $\psi_W$ i.e. $\star\psi_W=\frac{1}{\sqrt{3}}(\eta_1\eta_2+\eta_1\eta_3+\eta_2\eta_3)$, as before it is nonseparable and non-factorable in the weak sense. Namely, according to our observation concerning dual functions,
\begin{eqnarray}
w_{ij}(\star\psi_W|_{\eta_k=0})&=&0, \quad\quad \tilde{\mathcal{H}}_i=0,
\\
w_{ij}(\partial_k\psi_W)&=&-\frac{1}{3}
\end{eqnarray}
However, we should not be surprised to find that $\star\psi_W=\frac{\sqrt{3}}{2}(\psi_W)^2$.
\item[{\bf E2:}]
The GHZ state: $\psi_{GHZ}=\frac{1}{\sqrt{2}}(1+\eta_1\eta_2\eta_3)$. This function also is not factorable, even in a weak sense, because
\begin{eqnarray}
w_{ij}(\psi_{GHZ}|_{\eta_k=0})&=&0
, \quad\quad \tilde{\mathcal{H}}_k=\frac{1}{2}\\
w_{ij}(\partial_k\psi_{GHZ})&=&0,
\end{eqnarray}
 Let us note that $\psi_{GHZ}=\star\psi_{GHZ}$, and nontrivial contribution to the Wronskian $\eta$-function comes here only from the $\tilde{H}$ i.e. $w_{ij}(\psi_{GHZ})=\eta_k$.
\item[{\bf E3:}] In above examples functions cannot be expressed as a product of  other functions except one factor is trivial, now let us illustrate case when decomposition into product of functions is not the factorization in a weak sense. Namely, let $\psi=\frac{1}{2}(1+\eta_1\eta_2)(1+\eta_1\eta_3)(1+\eta_2\eta_3))=
    \frac{1}{2}(1+\eta_1\eta_2+\eta_1\eta_3+\eta_2\eta_3)$
\begin{eqnarray}
w_{ij}(\psi|_{\eta_k=0})&=& \frac{1}{4}, \quad\quad \tilde{\mathcal{H}}_k=0\\
w_{ij}(\partial_k\psi)&=& -\frac{1}{4}
\end{eqnarray}
and  even no weak factorization exists.
\item[{\bf E4:}] When we drop the first factor in above expression and normalize  new functions, then $\psi'=\frac{1}{\sqrt{3}}(1+\eta_1\eta_3+\eta_2\eta_3)$ and
\begin{eqnarray}\nonumber
&&w_{12}(\psi'|_{\eta_3=0})=0=w_{13}(\psi'|_{\eta_3=0}),\quad w_{23}(\psi'|_{\eta_1=0})=\frac{1}{3},\\\nonumber
&&\tilde{\mathcal{H}}_i=0\\
&&w_{12}(\partial_3\psi')=-\frac{1}{3}, \quad\quad w_{13}(\partial_2\psi')=0=w_{23}(\partial_1\psi'),
\end{eqnarray}
what here indicates weak factorization in $\eta_1$ and $\eta_2$ variables. One can express $\psi'$ as the product of two $n=2$ $GHZ$ states with common $\eta$-variable . Namely, $\psi'=\frac{1}{\sqrt{3}}(1+\eta_1\eta_3)(1+\eta_2\eta_3)$.
\item[{\bf E5:}]
 The example of the state $\psi=\frac{1}{\sqrt{3}}(\eta_2+\eta_3+\eta_1\eta_2\eta_3)$ shows that assumption of the vanishing $w_{ij}(\partial_k\psi)$ is important for the strong factorization. We have
\begin{eqnarray}\nonumber
w_{12}(\psi|_{\eta_3=0})&=&0=w_{13}(\psi|_{\eta_3=0}),\quad w_{23}(\psi|_{\eta_1=0})=-\frac{1}{3}, \quad\tilde{\mathcal{H}}_i=0\\
w_{12}(\partial_3\psi)&=&\frac{1}{3}=w_{13}(\partial_2\psi),\quad w_{23}(\partial_1\psi)=0,
\end{eqnarray}
Indeed, vanishing $w_{12}(\psi)$ and $w_{13}(\psi)$ indicate presence of weak
 factorizations: $\psi=\frac{1}{\sqrt{3}}(1+\eta_1\eta_2)(\eta_2+\eta_3)$ and $\psi=\frac{1}{\sqrt{3}}(1+\eta_1\eta_3)(\eta_2+\eta_3)$ respectively, but
  $w_{12}(\partial_3\psi)=w_{13}(\partial_2\psi)\neq 0$ show that the true separation of variables is not possible. Here we have product of rigged GHZ and Werner states.
\item[{\bf E6:}] Finally let $\psi$ be a factorable state (but not totaly factorable) e. g. $\psi=\frac{1}{2}(1+\eta_3+\eta_1\eta_2+\eta_1\eta_2\eta_3)$
\begin{eqnarray}
w_{12}(\psi|_{\eta_3=0})&=& \frac{1}{4},\quad\quad w_{12}(\partial_3\psi)= \frac{1}{4}, \quad\quad \tilde{\mathcal{H}_3}=\frac{1}{2}\\
w_{13}(\psi|_{\eta_2=0})&=& 0,\quad\quad   w_{13}(\partial_2\psi)= 0,\quad\quad \tilde{\mathcal{H}_2}=0\\
w_{23}(\psi|_{\eta_1=0})&=& 0,\quad\quad   w_{23}(\partial_1\psi)= 0, \quad\quad \tilde{\mathcal{H}_1}=0
\end{eqnarray}
In this case $w_{i3}(\psi)$ and $w_{i3}(\partial_j\psi)$ vanish, so dependence on $\eta_3$ can be factorized and indeed $\psi\sim(1+\eta_1\eta_2)(1+\eta_3)$ is a product of the GHZ-function and $e^{\eta_3}$.
\end{enumerate}
Our classification of strong factorization of the functions for the $n=3$ reveals the already known onion structure in the space of states discussed in \cite{miy, miyake-wad}. Firstly, we can distinguish three sets $B_i$ of mutually bipartite separable functions with nonempty common part of totaly separable functions. One can say that the $B_i$ form a rosette which is surrounded by  the set $B_W$ of nonseparable functions of W-type nonseparability (cf. above examples) and  of the set $B_{GHZ}$ of nonseparable functions of GHZ-type.
It is interesting, that if we count dimensions of our $\eta$-function spaces as follows: we take normalized functions; when function is separable we normalize each factor independently, then dimensions of above sets are the following
\begin{eqnarray}
&&\dim (B_i)=3+1=4\\
&&\dim(\cap_i B_i)=1+1+1=3\\
&&\dim (B_W)=2^3-1-1=6\\
&&\dim (B_{GHZ})=2^3-1=7
\end{eqnarray}
where for the $B_W$ we have taken into account that there vanishes body $F_0$
of the W-type function. These dimensions agree with the results obtained from the invariants theory \cite{miyake-wad}.
%
%
\subsection{Factorization of $F(\eta_1, \eta_2, \eta_3, \eta_4)$}
As before we shall consider Wronski $2\times2$ matrices to detect the possibility
of weak factorization. The determinants $w_{ij}(F(\eta_1, \eta_2, \eta_3, \eta_4))$ are functions of two $\eta$-variables
\begin{equation}
w_{ij}(F)=w_{ij}(F|_{\eta_k=\eta_l=0})+\tilde{\mathcal{H}}_k(F|_{\eta_l=0})\eta_k+
\tilde{\mathcal{H}}_l(F|_{\eta_k=0})\eta_l+\tilde{\mathcal{H}}_{kl}(F)\eta_k\eta_l,
\end{equation}
where e.g. $\tilde{\mathcal{H}}_{34}=\mathcal{H}+2F_3F_{124}+2F_4F_{123}-
2F_{13}F_{24}-2F_{14}F_{23}$ and $\mathcal{H}$ is the Cayley determinant given by Eq. (\ref{h4}). We use
notation $\pi_i(F)=F|_{\eta_i}$.
 Guided by the criterion for $n=3$ we can state the following separability
condition for $n=4$:\\[5mm]
\underline{\textbf{Strong factorization of $\eta$-functions  for n=4 (cases (i-j)(k-l)
and (i-j-k)(l)):}}\\[-1mm]
\begin{itemize}
\item[\textbf{(i-j)(k-l)}]
 When the function
$F(\eta_1, \eta_2,\eta_3,\eta_4)$ can be decomposed into the product of functions depending on separated pairs of variables $F=G(\eta_{i_1}, \eta_{i_2})\tilde{G}(\eta_{j_1},\eta_{j_2})$, $I=\{{i_1}, {i_2}\}$ and $J=\{{j_1},{j_2}\}$ $I\cup J=\{1,2,3,4\}$,
 where $i_1<i_2$, $j_1<j_2$ then,
    \begin{equation}
    w_{i_k j_l}(F)=0, \quad\quad w_{i_k j_l}(\partial_{i_{k'}} F)=0, \quad\quad w_{i_k j_l}(\partial_{j_{l'}} F)=0, \quad\quad w_{i_k j_l}(\partial_{i_{k'}}\partial_{j_{l'}} F)=0,
    \end{equation}
where $i_k\neq i_{k'}\in I$ and $j_l\neq j_{l'}\in J$.
\item[\textbf{ (i-j-k)(l)}]
 When a function $F(\eta_1, \eta_2,\eta_3,\eta_4)$ has decomposition into the product
    $F=G(\eta_{i_1}, \eta_{i_2}, \eta_{i_3})\tilde{G}(\eta_j)$ then,
    \begin{equation}
    w_{i_k j}(F)=0, \quad\quad w_{i_k j}(\partial_{i_{k'}} F)=0, \quad\quad w_{i_k j}(\partial_{i_{k''}} F)=0,\quad\quad
    w_{i_k j}(\partial_{i_{k'} i_{k''}} F)=0,
    \end{equation}
\end{itemize}
where $J=\{i_1, i_2, i_3\}$, $J=\{j\}$ and as before $I\cup J=\{1,2,3,4\}$;
$i_k\neq i_k'\neq i_{k''}$ .
The proof is given in the   \ref{sf4}. For functions with nonvanishing body there exists logarithm $ln(1+s(F))$ and in such a case above conditions are equivalent to the statements that either
\begin{equation}
\partial_{i_k}\partial_{j_l}ln(1+s(F))=0,\quad k,l=1,2
\end{equation}
for factorization $(i-j)(k-l)$
or
\begin{equation}
\partial_{i_k}\partial_{j}ln(1+s(F))=0, \quad k,=1,2,3
\end{equation}
for factorizations $(i-j-k)(l)$. We just get special cases of criterion  considered in Ref. \cite{mandi}. Let us stress that present criterions written directly in terms of Wronskians of $\eta$-function $F$ are more general and apply to all $\eta$-functions.

While for the $n=3$  we do not  have well defined  $3\times 3$ Wronski matrices,
in the present case we can introduce nontrivial $4\times 4$ ones in the
following way
\begin{equation}\label{L-mtx}
\mathcal{L}_{ij}
=\left(
\begin{array}{cc}
W_{ij}(F)& W_{ij}(\partial_k F)\\
W_{ij}(\partial_l F)&W_{ij}(\partial_l\partial_k F)
\end{array}
\right),
\end{equation}
where $(ij)$ and $(lk)$ are ordered indices, hence $i<j$, $l<k$. Firstly, let us
consider the body of these matrices, which can be written as conventional ones with the complex number's entries, in the following form
\begin{equation}\label{body-L-mtx}
L_{ij}
=\left(
\begin{array}{cc}
W_{ij}(\pi_l\pi_kF)& W_{ij}(\pi_l\partial_k F)\\
W_{ij}(\pi_k\partial_l F)&W_{ij}(\partial_l\partial_k F)
\end{array}
\right)
\end{equation}
It turns out that they are related to the known in the literature
matrices obtained from the invariants theory, namely using results  of Ref.(\cite{lu-th-67}) we get
\begin{equation}
det L_{12}=N, \quad\quad det L_{14}=M, \quad\quad det L^{PT}_{13}=L,
\end{equation}
where $L^{PT}_{13}$ means partial transposition of the matrix $L_{13}$
 \begin{equation}
L^{PT}_{13}
=\left(
\begin{array}{cc}
W_{13}(\pi_2\pi_4F)& W_{13}(\pi_4\partial_2 F)\\
W_{13}(\pi_2\partial_4 F)&W_{13}(\partial_2\partial_4 F)
\end{array}
\right).
\end{equation}
The presence of partial transposition is important here, because invariants $L$, $M$, $N$ satisfy the relation \cite{lu-th-67}
\begin{equation}\label{LMN}
L+N+M=0
\end{equation}
Such relation for Wronski matrices (\ref{L-mtx}) would be troublesome. As it is easy to see, the $\eta$-function valued determinants of $\mathcal{L}_{ij}$ give information about possibility of factorization and we are interested in all matrices $\mathcal{L}_{ij}$, however for the complementary pairs of indices we get the same values of the determinant, so is enough to consider e.g.: $L_{12}$, $L_{13}$, $L_{14}$.
While above conditions of separability for $n=4$ are invertible, we can get much weaker ones using determinants of $4\times 4$ matrices: $\mathcal{L}_{ij}$.
It turns out that, a necessary condition for $F(\eta_1,\eta_2,\eta_3,\eta_4)$ to be factorable into the product $G(\eta_{i_1}, \eta_{i_2}, \eta_{i_3})\tilde{G}(\eta_j)$ is the following
\begin{equation}
det \mathcal{L}_{i_k j}=0, \quad\quad k=1,2,3
\end{equation}
When $F$ is factorable into a product $F=G(\eta_{i_1}, \eta_{i_2})\tilde{G}(\eta_{j_1},\eta_{j_2})$ then
\begin{equation}
det \mathcal{L}^{PT}_{i_k j_l}=0, \quad\quad k,l=1,2.
\end{equation}
Let us note that for this type of separation of variables we take partial transposition of $\mathcal{L}_{ij}$.
To illustrate how above criterions differentiate various states let us consider some examples:
\begin{itemize}
\item[{\bf{E1:}}]For the $n=4$ GHZ-state $\psi_{GHZ}=\frac{1}{\sqrt{2}}(1+\eta_1\eta_2\eta_3\eta_4)$ we have
    \begin{eqnarray}
    &&w_{ij}(\psi_{GHZ})=\frac{1}{2}\eta_l\eta_k,\quad \mbox{note that:}\,
    b(w_{ij}(\psi_{GHZ}))=0\\
    &&w_{ij}(\partial_l\psi_{GHZ})=0\\
    &&w_{ij}(\partial_k\partial_l\psi_{GHZ})=0
    \end{eqnarray}
    and $ \mathcal{H}(\psi_{GHZ})=\frac{1}{2}$,  but $det\mathcal{L}_{ij}=0$ and also $M=N=L=0$. The Wronskians $w_{ij}$ are sensitive enough to detect nonfactorability, and what is important, not only the body of $w_{ij}$ is important indeed.
\item[{\bf{E2:}}] Similar result, we obtain for the $n=4$ W-state $\frac{1}{2}(\eta_1+\eta_2+\eta_3+\eta_4)$.  Here $\mathcal{H}(\psi_W)=0$, and
    \begin{eqnarray}
    &&w_{ij}(\psi_{W})=-\frac{1}{4}\\
    &&w_{ij}(\partial_l\psi_{W})=0\\
    &&w_{ij}(\partial_k\partial_l\psi_{W})=0.
    \end{eqnarray}
    This time, the nonzero content of the $w_{ij}$ is contained in the body of this Wronskian. As before: $det\mathcal{L}_{ij}=0$ and $M=N=L=0$. In conclusion we see, that here also the week criterion fails to detect essential nonseparability of the Werner state.
\item[{\bf{E3:}}] Present example shows that, the week criterion can be useful. Let us consider the following state:
    $\psi=\frac{1}{2}(\eta_3+\eta_4+\eta_1\eta_2\eta_3+\eta_1\eta_2\eta_4)$.
We have that $det \mathcal{L}_{ik}^{PT}=0=det \mathcal{L}_{ik}$ for $i=1,2$; $k=3,4$. But $det\mathcal{L}_{12}=det\mathcal{L}_{34}=\frac{1}{16}e^{\eta_1\eta_2}\neq 0$.
This signals actual factorability of this state, as $\psi=\psi_{GHZ}^{(12)}\psi_W^{(34)}$ and remaining nonfactorability inside pairs of variables (1-2) and (3-4).
\end{itemize}
%

%
%
\section{Entanglement monotones and flavors of entanglement of $\eta$-functions}
The task of finding a good definition of proper, sensitive and operational entanglement monotone is not trivial. Despite that there are accepted monotones for $n=2,3,4$ \cite{ver-deh-dem-ver} there is still
effort to modify them to suit special needs. Here we want to analyze known entanglement monotones in terms of $\eta$-functions and relate them to the criteria of the non/factorization of these functions. What is remarkable, that the entanglement monotones for pure states can be expressed in terms of Wronskians
of the $\eta$-functions representing relevant state. Such representation of the entanglement monotones gives an insight onto the flavor of entanglement of particular family of states.

In our discussion we shall follow the number of qubits and focus only on the pure state entanglement of bipartite systems.
\subsection{n=2}
The well known entanglement monotone for this case, the concurrence can be expressed
using the defined above Wronskian
\begin{equation}
\mathcal{C}(F(\eta_1, \eta_2))=2|w_{12}(F(\eta_1,\eta_2))|,\quad\quad
<F, F>=1,
\end{equation}
where the scalar product is defined in $\mathcal{N}$-Hilbert space $\mathcal{F}_0(\vec{\eta})$ and in components it takes the form
$<F, F>=|F_0|^2+|F_0|^2+|F_1|^2+|F_2|^2+|F_{12}|^2$. Using the notion of the comb \cite{ost-sie} and antilinear mapping $F\mapsto F^c$, where $F^c=(\sigma^y\otimes\sigma^y)\bar{F}$ one can express concurrence as
\begin{equation}
\mathcal{C}(F)=|<F^c,F>|=2|F_0F_{12}-F_1F_2|.
\end{equation}
The operator $\sigma^y\otimes\sigma^y$ is realized in the $\mathcal{F}_0(\vec{\eta})$ in the following form
\begin{equation}
\sigma^y\otimes\sigma^y=-(\partial_1\partial_2+\eta_1\eta_2
-\eta_2\partial_1-\eta_1\partial_2)
\end{equation}
It is interesting that by taking modulus of scalar product of the entries of defined above Wronski matrix i.e.
\begin{equation}
\mathcal{V}_i=2|<\partial_i F, F>|
\end{equation}
and analogously defining
\begin{equation}
\mathcal{P}_i=|<F,J_i(F)>|,
\end{equation}
where $J_1(F(\eta_1, \eta_2))=F(-\eta_1, \eta_2)$ and $J_2(F(\eta_1, \eta_2))=F(\eta_1, -\eta_2)$,
we reproduce so called complementarity relations
\begin{equation}
\mathcal{C}^2(F)+\mathcal{V}_i^2(F) +\mathcal{P}_i^2(F)=<F,F>^2=1, \quad i=1,2
\end{equation}
Using  projections $\pi_{k|0}$ and $\pi_{k|1}$ we can express known in the literature parameters $w$, $z$, $\zeta$ \cite{levay-geo} by the $\eta$-scalar product and $\mathcal{D}_1$ of a $\eta$-function $F$
\begin{eqnarray}
w&=&\mathcal{D}_1(F(\eta_1),\,\tilde{F}(\eta_1))=\int F(\eta_1)\wedge\tilde{F}(\eta_1)d\eta_1\\\nonumber
&=& \int F(\eta_2)\wedge\tilde{F}(\eta_2)d\eta_2=\mathcal{D}_1(F(\eta_2),\,\tilde{F}(\eta_2))\\
\zeta&=&<F(\eta_1),\,\tilde{F}(\eta_1)>=
\int\bar{F}(\eta_1)\tilde{F}(\eta_1)e^{\bar{\eta}_1\eta_1}d\bar{\eta}_1d\eta_1\\
z&=&<F(\eta_2),\,\tilde{F}(\eta_2)>=
\int\bar{F}(\eta_2)\tilde{F}(\eta_2)e^{\bar{\eta}_2\eta_2}d\bar{\eta}_2d\eta_2
\end{eqnarray}
and we have that $\mathcal{C}=2|w|$, $\mathcal{V}_1=2|z|$, $\mathcal{V}_2=2|\zeta|$.
\subsection{n=3}
The 3-tangle \cite{w, ckw} is the principal entanglement monotone used to detect and measure the "degree" of non-separability of 3-qubit systems. It makes use of the hyperdeterminant known in invariants theory for a long time
\begin{equation}
\tau_{123}=4|Det(F)|
\end{equation}
It is sensitive enough to detect the
GHZ state entanglement, which is maximal, but it neglects the entangled character
of the Werner state. In the cited above paper, Coffman Kundu and Wooters derived
relation for the 3-tange $\tau_{123}$ and mutual concurrences of bipartite systems of three
qubits (we number qubits instead labeling them by the letter) in the following
form
\begin{equation}
\mathcal{C}_{1(23)}^2=\mathcal{C}_{12}^2+\mathcal{C}_{13}^2+\tau_{123}
\end{equation}
Using this equation, one can define entanglement monotone symmetric with respect qubit indices, averaged over possible configurations of qubits.
Namely,
\begin{equation}
Q(F)=\frac{1}{3}(\mathcal{C}_{1(23)}^2+\mathcal{C}_{2(13)}^2+\mathcal{C}_{3(12)}^2)
=\frac{2}{3}(\mathcal{C}_{12}^2+\mathcal{C}_{13}^2+\mathcal{C}_{23}^2)
+\tau_{123}
\end{equation}
Let us note that the $Q$ is the $n=3$ realization of global entanglement measure introduced by Meyer and Wallach \cite{mey-wall-jmp} for arbitrary $n$ (cf. also \cite{bren}for other form of this function).

It can be seen that (cf.   \ref{Det-wron} for the proof)
\begin{equation}\label{det-wron}
Det(F)=\frac{1}{3}\sum_k(\tilde{\mathcal{H}}_k^2-4w_{ij}(F|_{\eta_k=0})
w_{ij}(\partial_k F), \quad\quad i\neq j \neq k
\end{equation}
Using $\eta$-realization of $\sigma$ matrices and the scalar product in
$\mathcal{N}$- Hilbert, one gets that
\begin{equation}
<\sigma_2^{ij}\bar{F},F>=-2\left(w_{ij}(F|_{\eta_1=0})+w_{ij}(\partial_kF)\right),
\end{equation}
where bar denotes complex conjugation and $\sigma_2^{ij}$ is a tensor product of $\mathbb{I}$ and $\sigma_2$ matrices on $i^{th}$ and $j^{th}$ positions e.g.
$\sigma_2^{23}=\mathbb{I}\otimes\sigma_2\otimes\sigma_2$.
Let us consider the following monotone
\begin{equation}
\mu(F)=\frac{2}{3}\left(\sum_{i<j} |<\sigma_2^{ij}\bar{F},F>|^2+\tau_{123}\right)
\end{equation}
\begin{equation}
\mu(F)=\frac{4}{3}\left(2\sum_k|w_{ij}(F|_{\eta_k=0})+w_{ij}(\partial_kF)|^2
+|\sum_k(\tilde{\mathcal{H}}_k^2-4w_{ij}(F|_{\eta_k=0})w_{ij}(\partial_kF)|\right)
\end{equation}
With the use of relation (\ref{tr-w}) the $\mu(F)$ can be expressed solely in terms of determinants and traces of the Wronski matrices $W_{ij}(F)$, $W_{ij}(\partial_k F)$ namely,
\begin{eqnarray}
\mu(F)&=&\frac{4}{3}(2\sum_k|w_{ij}(F|_{\eta_k=0})+w_{ij}(\partial_kF)|^2
+|\sum_k((tr W_{ij}(F|_{\eta_k=0})\\&&\cdot tr W_{ij}(\partial_k F)-
-tr( W_{ij}(F|_{\eta_k=0}) W_{ij}(\partial_k F)))^2
-4w_{ij}(F|_{\eta_k=0})w_{ij}(\partial_kF)|).\nonumber
\end{eqnarray}
Above representation shows explicitly how entanglement and factorability properties are intertwined.

From our previous discussion on Wronskians and $\tilde{\mathcal{H}}_i$ it follows that duality transformation preserves the value of $\mu$ i.e. $\mu(F)=\mu(\star F)$.
Let us test its behavior on some states. It will be instructive to come back to the functions considered in examples illustrating types of factorability of  $\eta$-functions in Sec \ref{fac-3}.
Now the normalization of $\eta$-function is important, to have fixed scale of values for $\mu$.\\
\textbf{\underline{Examples:}}
\begin{itemize}
\item[{\bf E1:}] $\mu(\frac{1}{\sqrt{3}}(\eta_1+\eta_2+\eta_3))=\frac{8}{9}$\\
The $\psi_W$ state contributes to the value of $\mu$ only by the relative 2-qubit entanglement, the 3-tangle $\tau_{123}$ vanishes. Analogously situation is for $\star\psi_W$;
 $\mu(\frac{1}{\sqrt{3}}(\eta_1\eta_2+\eta_1\eta_3+\eta_2\eta_3))=\frac{8}{9}$
\item[{\bf E2:}] The GHZ state cotributes to the value of $\mu$ only through  the 3-tangle
$\mu(\psi_{GHZ}=\frac{1}{\sqrt{2}}(1+\eta_1\eta_2\eta_3))=1$
\item[{\bf E3:}] It is interesting to compare behavior of $\mu$ on states of the form $\sim 1+\star\psi_W$ and $\sim 1+\psi_W$. For the first one we get $\mu(\frac{1}{2}(1+\eta_1\eta_2+\eta_1\eta_3+\eta_2\eta_3))=1$ with contribution solely from 3-tangle, but for the second state we get only contribution from the relative 2-qubit entanglement and
    $\mu(\frac{1}{2}(1+\eta_1+\eta_2+\eta_3))=\frac{3}{4}$.
\item[{\bf E4:}] The function in this example and the one in the \textbf{E5} give the same value of $\mu$ with almost the same mechanism, in both cases 3-tangle vanishes;
      $\mu(\frac{1}{\sqrt{3}}(1+\eta_1\eta_3+\eta_2\eta_3))=\frac{8}{9}$.
\item[{\bf E5:}] $\mu(\frac{1}{\sqrt{3}}(\eta_2+\eta_3+\eta_1\eta_2\eta_3))=\frac{8}{9}$
\item[{\bf E6:}] Here we have nonvanishing $\tilde{\mathcal{H}}_3$ and 3-tangle vanishes due to some cancelations of terms, again nonzero contribution to $\mu$ comes from the 2-qubit subsystem entanglement; $\mu(\psi=\frac{1}{2}(1+\eta_3+\eta_1\eta_2+\eta_1\eta_2\eta_3))=\frac{1}{6}$.
\item[Note:] Taking $\psi_W^{(12)}=\frac{1}{\sqrt{2}}(\eta_1+\eta_2)$ and $\psi_{GHZ}^{(12)}=\frac{1}{\sqrt{2}}(1+\eta_1\eta_2)$ as the 3-qubit states with trivially factorized third qubit we get that $\mu(\psi_W^{(12)})=\frac{2}{3}$ and $\mu(\psi_{GHZ}^{(12)})=\frac{2}{3}$.
\end{itemize}
\subsection{n=4}
\subsubsection{Representatives of equivalence classes of entangled $\eta$-wavefunctions}
Here we shall use results of the approach to entanglement based SLOCC
classification. The problem of equivalence of the non-normalized pure states
under
permutations and the SLOCC transformations having determinant equal one was solved
with use the invariants theory \cite{ver-deh-dem-ver, chte-dok, lu-th-67}. The representatives
were found originally by \cite{ver-deh-dem-ver} but we shall follow the modified  form obtained in \cite{chte-dok}. In terms of elementary $\eta$-functions these nine invariants take the following form (for convenience of the reader natural form these functions is given in   \ref{psi-4})
%
\begin{eqnarray}
\Psi_1&=&\frac{a+d}{2}e^{\vec{\eta}}+\frac{a-d}{2}(\cos(\eta_1-\eta_2)-
\cos(\eta_3+\eta_4))
+\frac{b+c}{2}(\cos(\eta_1-\eta_3)\\\nonumber
&&-\cos(\eta_2+\eta_4))
+\frac{b-c}{2}(\cos(\eta_1-\eta_4)-\cos(\eta_2+\eta_3))
\\\nonumber
%
\Psi_2&=&\frac{a+c-i}{2}e^{\vec{\eta}}+\frac{a-c+i}{2}(\cos(\eta_1-\eta_2)
-cos(\eta_3+\eta_4))
+\frac{b+c+i}{2}(\cos(\eta_1-\eta_3)\\\nonumber
&&-\cos(\eta_2+\eta_4))
+\frac{b-c-i}{2}(\cos(\eta_1-\eta_4)-\cos(\eta_2+\eta_3))
+\frac{i}{2}(\sin(\eta_1+ \eta_2+\eta_4)\\
&&-\sin(\eta_2+\eta_3+\eta_4)+
+\sin(\eta_1+\eta_3+\eta_4)-\sin(\eta_1+ \eta_2+ \eta_3))
\\\nonumber
%
\Psi_3&=&\frac{a}{2}e^{\eta_1\eta_2+\eta_3\eta_4}+\frac{b+1}{2}(\cos(\eta_1-\eta_3)
-\cos(\eta_2+\eta_4))
+\frac{b-1}{2}(\cos(\eta_1-\eta_4)\\
&&-\cos(\eta_2+\eta_3))+
+\frac{1}{2}(\sin(\eta_1+\eta_2+\eta_3)-\sin(\eta_1+\eta_2+\eta_4))
\\\nonumber
%
%
\Psi_4&=&\frac{a+b}{2}e^{\vec{\eta}}+b(\cos(\eta_1-\eta_3)-\cos(\eta_2+\eta_4))
+i(\cos(\eta_2-\eta_3)-\cos(\eta_1+\eta_4)\\\nonumber
&&-2\sin\eta_2\sin\eta_3)+
+\frac{a-b}{2}(\cos(\eta_1-\eta_2)-\cos(\eta_3+\eta_4))
+\frac{1}{2}(\sin(\eta_1-\eta_2-\eta_3+\eta_4)\\
&&+\sin(\eta_1+\eta_4)-\sin(\eta_2+ \eta_3))
\end{eqnarray}
%
\begin{eqnarray}
%
%
\Psi_5&=&\frac{a}{2}e^{\eta_1\eta_2+\eta_3\eta_4}
-2i(\sin\eta_2\cos(\eta_1+\eta_3)+\cos(\eta_1+\eta_4)-1)
\\\nonumber
%
%
\Psi_6&=&\frac{a+i}{2}e^{\eta_1\eta_2+\eta_3\eta_4}+\frac{a+i+1}{2}(\cos(\eta_1-\eta_3)
-\cos(\eta_2+\eta_4)
+\frac{a-i-1}{2}(\cos(\eta_2-\eta_3)-\\\nonumber
&&-\cos(\eta_1+\eta_4))+
\frac{i}{2}(\sin(\eta_1+\eta_3+\eta_4)-\sin(\eta_1+\eta_2+\eta_4)+
\sin(\eta_2+\eta_3+\eta_4)-\\
&&-\sin(\eta_1+\eta_2+\eta_3))
+\frac{1}{2}(\sin(\eta_1+\eta_2+\eta_3)-\sin(\eta_1+\eta_2+\eta_4))
\\\nonumber
%
%
\Psi_7&=&\sin\eta_1\sin(\eta_1+\eta_3+\eta_4)+\sin\eta_2\sin(\eta_1-\eta_3+\eta_4)
+i(\sin(\eta_2+\eta_3+\eta_4)\\
&&+\sin(\eta_1-\eta_2-\eta_4)-
-\sin(\eta_1+\eta_2+\eta_3))
\\\nonumber
%
%
\Psi_8&=&\frac{1}{2}(\cos(i(\eta_1+\eta_2+\eta_3+\eta_4))-
\sin(\eta_1+\eta_2-\eta_3-\eta_4)+\sin(\eta_1+\eta_2)-\sin(\eta_3+\eta_4)+\\\nonumber
&&+\sin(\eta_1-\eta_3)+\sin(\eta_2-\eta_4))+
\frac{i}{2}(\cos(\eta_1+\eta_2+\eta_3+\eta_4)+
\cos(\eta_1+\eta_3+\eta_4)-\\
&&-\cos(\eta_2-\eta_3)
-2\sin(\eta_1+\eta_3-\eta_4)+
+\sin(\eta_1-\eta_2+\eta_3)-\sin(\eta_1-\eta_2+\eta_4))
\\
%
%
\Psi_9&=&\frac{1}{2}e^{\eta_1+\eta_2}\cos(\eta_3-\eta_4))
+\frac{i}{2}e^{\eta_1-\eta_2}\sin(\eta_3+\eta_4))
\end{eqnarray}
where $\vec{\eta}=\vec{\eta}_4=\eta_1\eta_2\eta_3\eta_4$. It turns out that
$\Psi_k$ for $k=1,2,3,6$ are selfdual i.e. $\star(\Psi_k)=\Psi_k$. Let us observe that representation by means of the elementary $\eta$ function is fluent in the sense that we have various identities for them, which can be used to change actual form of $\Psi_i$. As noted by by Verstraete et all \cite{ver-deh-dem} the $G_{abcd}$ state (here it is $\Psi_1$ function) is of generic type (any pure state of $4$-qubits can be transformed into it) and with maximal $4$-partite entanglement. It is remarkable that using $\eta$-functions it is easy to note that
\begin{equation}
G_{abcd}=\Psi_1=\frac{a}{2}e^{\eta_1\eta_2}e^{\eta_3\eta_4}
+\frac{d}{2}e^{-\eta_1\eta_2}e^{-\eta_3\eta_4}+
\frac{b}{2}(\eta_1+\eta_2)(\eta_3+\eta_4)+
\frac{c}{2}(\eta_1-\eta_2)(\eta_3-\eta_4)
\end{equation}
what means a sum of "diagonal" products of the GHZ and Werner states respectively of two-qubit subsystems of our total 4-qubit system. Namely,
\begin{equation}
G_{abcd}=\Psi_1=\frac{a}{2}\psi_{GHZ+}^{(12)}\psi_{GHZ+}^{(34)}+
\frac{d}{2}\psi_{GHZ-}^{(12)}\psi_{GHZ-}^{(34)}+
\frac{b}{2}\psi_{W+}^{(12)}\psi_{W+}^{(34)}+
\frac{c}{2}\psi_{W-}^{(12)}\psi_{W-}^{(34)}
\end{equation}
Above factorized form shows the flavor of  entanglement of the $\Psi_1$ function.
\subsubsection{Entanglement monotones based on polynomial invariants}
As it is well studied in the literature one of the main invariants in this case is the Cayley determinant.
Another set of invariants consists of the determinants $D_{uv}=det(B_{uv})$ \cite{lu-th-67}. Luque and Thibon give explicit form of them. It appears that in terms of the $\eta$-functions they are related to the following Wronski matrices
\begin{equation}\label{B-mtx}
B_{ij}
=\left(
\begin{array}{ccc}
\mathcal{H}(\pi_i\pi_j F)&\tilde{\mathcal{H}}_{j}(\pi_i F) & \mathcal{H}(\pi_i\partial_j F)\\
\tilde{\mathcal{H}}_i(\pi_j F)& \tilde{\mathcal{H}}_{ij}(F)&\tilde{\mathcal{H}}_i(\partial_j F)\\
\tilde{\mathcal{H}}(\partial_i\pi_j F)&\tilde{\mathcal{H}}_j(\partial_i F)& \mathcal{H}(\partial_i\partial_j F)
\end{array}
\right),
\end{equation}
where the $\mathcal{H}$, $\tilde{\mathcal{H}}_i$ and $\tilde{\mathcal{H}}_{ij}$ are taken in appropriate form for $\eta$-functions with $n=2, 3, 4$ variables.  Taking determinants for pairs $(ij)=(12), (13), (14)$ we get invariants which are equal to $D_{xy}=det(B_{xy})$, $D_{xz}=det(B_{xz})$ and $D_{xt}=det(B_{xt})$ in the notation of Ref.\cite{lu-th-67}. From the identity (\ref{LMN}) it follows that
\begin{equation}
L^2+M^2+N^2=-2(MN+NL+ML).
\end{equation}
To get symmetric (permutation invariant) monotones let us use the so called Schl\"{a}fli basis $\{\mathcal{H}, W,\Sigma, \Pi\}$, where $W=D_{xy}+D_{xz}+D_{xt}$, $\Sigma=L^2+M^2+N^2$ and $\Pi=(L-M)(M-N)(N-L)$. Osterloh and Siewert introduced for $n=4$ the entanglement monotones $|\mathcal{F}_i|$; $i=1,2,3,4,5$, which can be expressed in the Schl\"afli basis as follows (we modify slightly the form of relations from  \cite{ost-sie, ren-zho})
\begin{eqnarray}
\mathcal{F}_1&=&8(4W-\mathcal{H}^3)\\
\mathcal{F}_2&=&16(\mathcal{H}^4-4\mathcal{H}W-4(\mathcal{H}D_{xt}+4LM))\\
\mathcal{F}_3&=&32(\mathcal{H}^6-24\mathcal{H}^2\Sigma-64\Pi)\\
\mathcal{F}_4&=&16(\mathcal{H}^4-4\mathcal{H}W-4(\mathcal{H}D_{xz}+4LN))\\
\mathcal{F}_5&=&16(\mathcal{H}^4-4\mathcal{H}W-4(\mathcal{H}D_{xy}+4MN))
\end{eqnarray}
Therefore the $|\mathcal{F}_i|$ for $i=2,4,5$ are not symmetric, but can be combined into the symmetric entanglement monotone. As proposed in \cite{ren-zho} one can take a combination of them to get symmetric entanglement monotone
\begin{equation}
|\mathcal{F}_2'|=|\mathcal{F}_2+\mathcal{F}_4+\mathcal{F}_5|
=16|3\mathcal{H}^4-16\mathcal{H}W+8\Sigma|
\end{equation}
Let us check how such defined symmetric entanglement monotone behaves on the following nontrivially entangled states:
\begin{itemize}
\item the GHZ-state: $\psi_{GHZ}=\frac{1}{\sqrt{2}}(1+\eta_1\eta_2\eta_3\eta_4)$, $\star\psi_{GHZ}=\psi_{GHZ}$
\item the Werner state: $\psi_W=\frac{1}{2}(\eta_1+\eta_2+\eta_3+\eta_4)$ and dual to it state
$\star\psi_W=\frac{1}{2}(\eta_1\eta_2\eta_3+\eta_1\eta_2\eta_4+\eta_1\eta_3\eta_4+\eta_2\eta_3\eta_4)$
\item the cluster Werner state:
$\psi_{CW}=\frac{1}{\sqrt{6}}(\eta_1 \eta_2+\eta_1\eta_3+\eta_1\eta_4+\eta_2\eta_3+
\eta_2\eta_4+\eta_3\eta_4)$. It is a selfdual state $\star\psi_{CW}=\psi_{CW}$
\item $\phi^{(ij)}$-family: $\phi_{\pm\pm}^{(ij)}=
\frac{\sqrt{2}}{2}(\psi_{W\pm}^{(ij)}+\eta_k\eta_l\psi_{GHZ\pm}^{(ij)})$,
 i.e. $\phi_{++}^{(ij)}$, $\phi_{--}^{(ij)}$, $\phi_{+-}^{(ij)}$, and $\phi_{-+}^{(ij)}$,
\item $\chi^{(ij)}$-family: $\chi_{\pm\pm}^{(ij)}=
\frac{\sqrt{2}}{2}(\psi_{GHZ\pm}^{(ij)}+\eta_k\eta_l\psi_{W\pm}^{(ij)})$, i.e. $\chi_{++}^{(ij)}$, $\chi_{--}^{(ij)}$, $\chi_{+-}^{(ij)}$, and $\chi_{-+}^{(ij)}$,
\item $\psi^{(ij)}$-family:
$\psi_{\pm\mp}^{(ij)}=
\frac{\sqrt{2}}{2}(\psi_{GHZ\pm}^{(ij)}+\eta_k\eta_l\psi_{GHZ\mp}^{(ij)})$, i.e. $\psi_{+-}^{(ij)}$, $\psi_{-+}^{(ij)}$
\item $\lambda^{(ij)}$-family:
$\lambda_{\pm\mp}^{(ij)}=
\frac{\sqrt{2}}{2}(\psi_{GHZ\pm}^{(ij)}+\eta_k\eta_l\psi_{GHZ\mp}^{(ij)})$, i.e. $\lambda_{+-}^{(ij)}$, $\lambda_{-+}^{(ij)}$.
\item $\Phi_{A\pm}^{(jk,l)}$-family:
$\Phi_{A\pm}^{(jk,l)}=\frac{\sqrt{3}}{2}(\psi_{GHZ\pm}+\eta_j\psi_W^{(kl)}+\eta_k\psi_W^{(il)})$
    e.g.
    $\Phi_{A\pm}^{(23,4)}=\frac{\sqrt{3}}{2}(\psi_{GHZ\pm}+\eta_2\psi_W^{(34)}+\eta_3\psi_W^{(14)})$
\item $\tilde{\Phi}_{\pm}$ -states: $\tilde{\Phi}_{\pm}=\frac{1}{2}(\psi_{GHZ\pm}+\sqrt{3}\psi_{CW})$,
\end{itemize}
where we take strictly ordered multi-indices $(ij)$, or $(ij,k)$ .
Above families of states are spanned on pairs of the 2-qubit GHZ-states and W-states of the respective subsystems, or as in the case of $\Phi_{A\pm}, \tilde{\Phi}_{\pm}$ on total $\psi_{GHZ}$ state and relevant $\psi_{(ij)}$, or $\psi_{CW}$ states. From the $\eta$-function point of view they are natural and interesting as vectors in the $\mathcal{N}$-Hilbert space. Some examples of them already appeared in the literature in the context of entanglement, but without reference to the decomposition we use e.g. three states belonging to the $\phi_{++}^{(ij)}$ family ($\phi_{++}^{(1,3)}$, $\phi_{++}^{(2,3)}$, $\phi_{++}^{(34)}$ ) were discussed in Ref. \cite{ren-zho} and the $\Phi_{A\pm}^{(23,4)}$ was used in the Ref. \cite{b-y-w}
It is fruitful to compare behavior of $|\mathcal{F}_2'|$ and $|\mathcal{F}_3|$ on above sets of states. We collect their values in the Table \ref{em-f2}, together with the values of invariants contributing to the $|\mathcal{F}_2'|$  and $|\mathcal{F}_3|$. Taking into account the way the final value is achieved one finds, that the cluster Werner state,   $\Phi_{A\pm}^{(jk,l)}$-family and $\tilde{\Phi}_{\pm}$-family of states are interesting, because they give contributions from different channels: $\mathcal{H}, W, \Sigma, \Pi$. The family  $\Phi_{A+}^{(ij,k)}$ is exceptional, in the sense that, despite nontrivial contributions from all invariants: $\mathcal{H}, W, \Sigma$ $\Pi$ both monotones give zero as the entanglement measure. Other states are detected, if at all, only by one of available invariants contributing to entanglement monotones. The entanglement of the Werner state (or dual Werner state $\star\psi_W$) is not detected at all. As we already discussed for the $\psi_W$, in that case one can get contributions directly from the Wronskian $w_{ij}(\psi_W)=-\frac{1}{4}$. It is worth noting that $|\mathcal{F}_2'|$ and $|\mathcal{F}_3|$ give similar information about entanglement of considered states. Simultaneous vanishing of their values for arbitrary state is not obvious from the definitions of these entanglement monotones, except a particular case when the Cayley determinant  $\mathcal{H}$ for a state vanishes and one of the determinants $M$, $N$ or $L$ is zero. Let, for example $M\neq 0$ and $L=0$, then $N=-M$, and we have that
\begin{eqnarray}
|\mathcal{F}_2'|&=& 2^8|M|^2,\\
|\mathcal{F}_3|&=&2^{12}|M|^3,
\end{eqnarray}
or equivalently
\begin{equation}
|\mathcal{F}_3|=2^4|M||\mathcal{F}_2'|.
\end{equation}
We meet such a case for the families of states: $\psi_{\pm\mp}^{(ij)}$, $\lambda_{\pm\mp}^{(ij)}$,
$\phi_{\pm\pm}^{(ij)}$, $\chi_{\pm\pm}^{(ij)}$, as well as for the states $\phi_{A-}^{(jk,l)}$.
%
\begin{table}
\caption{Values of the entanglement monotones $|\mathcal{F}_2'|$ and $|\mathcal{F}_3|$ on
selected
families of states.}\label{em-f2}
\begin{flushleft}
\begin{tabular}{cccccccccccc}
\hline\hline
 & $\psi_{GHZ}$ & $\psi_W$ & $\psi_{CW}$ & $\phi_{\pm\pm}^{(ij)}$ &$\chi_{\pm\pm}^{(ij)}$&  $\psi_{+-}^{(ij)}$ & $\lambda_{+-}^{(ij)}$
 &$\phi_{A+}^{(jk,l)}$&$\phi_{A-}^{(jk,l)}$&$\tilde{\Phi}_{A+}$&$\tilde{\Phi}_{A-}$\\[0.5mm]
\hline\hline
$|\mathcal{F}_2'|$ & 3 & 0 & $\frac{11}{9}$ & 1&1&1&1&0&$(\frac{2}{3})^4$&3&$(\frac{3}{4)}^2$\\
$|\mathcal{F}_3|$ & $\frac{1}{2}$ & 0 & $\frac{1}{2}$ & 1&1&1&1&0&$(\frac{2}{3})^6$&$\frac{1}{2}$&$(\frac{1}{2})^7$\\
\hline
$\mathcal{H}$&$\frac{1}{2}$&0&$\frac{1}{2}$&0&0&0&0&$\frac{1}{3}$
&0&$\frac{1}{2}$&$\frac{1}{4}$\\
W&0&0&$\frac{1}{72}$&0&0&0&0&$2(\frac{1}{6})^3$&$2(\frac{1}{6})^3$&0&$3(\frac{1}{2})^{8}$\\
$\Sigma$&0&0&0&$\frac{1}{2^7}$&$\frac{1}{2^7}$&$\frac{1}{2^7}$&$\frac{1}{2^7}$&
$2(\frac{1}{6})^4$&$2(\frac{1}{6})^4$&0&0\\
$\Pi$&0&0&0&$-(\frac{1}{2})^{11}$&$(\frac{1}{2})^{11}$&$(\frac{1}{2})^{11}$
&$(\frac{1}{2})^{11}$&$-2(\frac{1}{6})^{6}$&$2(\frac{1}{6})^{6}$&0&0\\[1mm]
\hline\hline
\end{tabular}
\end{flushleft}
\end{table}
%
\section{Conclusions}\label{concl}
Description of the entanglement is a complex task, therefore any formulation which allows to see the problem from a
new perspective is welcomed and may help in getting desired answer. The
 analysis of the qubit systems presented in this paper makes use of the new tool suitable for description of two level boson-like object - the nilpotent commuting variables. In spite of the nilpotency, the property which such variables share with the Grassmannian ones (used in supersymmetric theories), the nilpotent commuting
variables yield theory which is different then one known from supersymmetry. There,  fermionic systems are described by means of the anticommuting, hence nilpotent
variables with the use of the well developed supermathematics. Here, the differential
calculus is not the supercalculus, but the new one, with modified Leibniz rule. For qubits, we do not want to use the anticommutation relations suitable for fermions, but qubit commutation relations which are, on the other hand, not bosonic ones, but define parafermions. Qubit has two
dimensional representations and commuting nilpotent variables automatically provide
correct properties of the single qubit and multiqubit systems. In analogy to the
(super)symmetric theories there exists the classical mechanics for systems
described by commuting nilpotent variables \cite{amf-ncm}. It is worth noting that, nilpotent variables emerge also in quantum field theory of complex systems. Supporting the idea that the nilpotent commuting variables give natural language for description of two level systems, but composed one - not fermionic.

In the present paper there is developed further the formalism which can be used for the description of qubits. This formalism seems to be effective, practical and easy to interpret. Firstly, in this context nilpotent commuting variables were used by Mandilara at al. \cite{mandi, mandi2}. On the other hand, the present author developed the formalism suitable to describe classical systems with nilpotent commuting coordinates, in analogy to psedomechanical systems described by anticommuting variables. The peculiarities of appropriate differential calculus and an analog of the variational calculus were studied \cite{amf-ncm}. Hence, by now, various building blocks of the theory involving nilpotent commuting variables are at hand.
In this work we have introduced the $\mathcal{N}$-Hilbert space of $\eta$-wave
functions allowing to formulate the Schr\"{o}\-din\-ger description of quantum system including the analog of the Schr\"{o}\-din\-ger equation (cf. \cite{mandi}). It is remarkable that simple two qubit system, when not degenerate has the eigenvectors system consisting of Bloch vectors, and hence entangled. We have also addressed the question of hybrid, composed systems in analogy to the supersymmetric boson-fermion systems. As
specially interesting we find the qubit-fermion system. It is really
supersymmetric, in its simplest form it is composed of two, two level systems,
having superalgebra of charges and supersymmetry transformations. It exhibits the effect present also in the conventional quantum supersymmetric systems with a periodic potential, that the Witten index vanishes.

Many characteristic state vectors considered in literature in the context of
entanglement have very simple and natural representation as $\eta$-functions. They are just exponents, or trigonometric functions. In the present paper there are also  introduced and studied the symmetric $\eta$-polynomials, Hermite $\eta$-polynomials etc. The strong indication that $\eta$-functions are well tailored for representing  the qubit systems comes also from the description of entanglement. Invariants known in the literature for
various $n=2,3,4$ brought to the physics from the classical invariants theory,
frequently with complicated origins, here appear as simple expressions in Wronskians of
$\eta$-functions. The questions of flavors of pure state entanglement and factorization
within this formalism have natural, intuitive setting.

We hope that the results of the Mandilara et al. \cite{mandi} and results obtained in the present work show that the $\eta$-formalism is suitable and effective tool for the description of entanglement. We think  that it is something more then an ad hoc tool. The nilpotent classical mechanics \cite{amf-ncm} and nilpotent quantum mechanics as well as  field theory with nilpotent commuting variables \cite{palu-1, amf-nf} reveal reach structure and deserve further study, although they describe non-fundamental particles, but composite ones.
%
\begin{acknowledgements}
The author thanks cordially Andrzej Borowiec for many discussions and Jaromir Simsa for making available his works on factorizations of functions of conventional variables.
\end{acknowledgements}
%
\appendix{Permanent and Hafnian}\label{perm}
As it is known there are four important functions from the
set of square matrices into the ring or field over which matrices
are considered. Namely determinant and permanent, then Pfaffian
and Hafnian \cite{cai}. The permanent, like determinant is defined
for any square matrix, Pfaffian  for antisymmetric matrices and
the Hafnian  for symmetric ones, but in both cases $n=2k$. The
determinant and Pfaffian are sensitive to the parity of
permutation while permanent and Hafnian are not. This makes great
difference in properties of above objects and the level of
complication in computing them. The determinant/Pfaffian is very
universal object and more easy to handle then permanent/Hafnian,
its computational complexity is of polynomial type while
according to the Valiant's conjecture: there is no
polynomial-sized formula for the permanent \cite{val}. This is
important obstruction in calculating plane partitions in any
models using bipartite planar graphs. Such a questions are
extremely important
in e.g. dimer problem in statistical mechanics.\\
Let $A$ be an $n\times n$ matrix, the permanent of $A=(a_{ij})$ is
a sum over permutations $S_n$
\begin{equation}
per(A)=\sum_{\sigma\in S_n} \prod a_{i\sigma(i)},
\end{equation}
so $det$ is just a signed sum of the same type. Analogous relation
is for Pfaffian and Hafnian.\\
Let $A_{2n\times 2n}$ be a symmetric matrix and $\alpha$ be a
partition i.e. $\alpha\{(i_1,j_1),i_2,j_2),\dots, (i_n,j_n)\}$
(where $i_k<j_k$, $i_1<i_2<\dots<i_n$) and $\sigma$ be a
corresponding permutation
$$
\sigma=\left(
\begin{array}{llllcll}
1&2&3&4&\dots&2n-1&2n\\
i_1&j_1&i_2&j_2&\dots&i_n&j_n
\end{array}
\right)
$$
\begin{equation}
\alpha \longrightarrow A_{\alpha}=a_{i_1 j_1}a_{i_2 j_2}\dots
a_{i_n j_n}
\end{equation}
By Pfaffian one understands the signed sum
\begin{equation}
P\!f(A)=\sum_{\alpha\in S_{2n}} (-1)^{|\alpha|}A_{\alpha},
\end{equation}
and by Hafnian, $H\!f(A)$,
\begin{equation}
H\!f(A)=\sum_{\alpha\in S_{2n}} A_{\alpha},
\end{equation}

The Pfaffian satisfies, among others, two important identities
\cite{cai}
\begin{equation}
P\!f(A)^2=det(A),\quad P\!f(B^TAB)=det(B)P\!f(A).
\end{equation}
Analogous relations, in general, are not valid for the permanent
and Hafnian (cf. \cite{cai})
\appendix{Index conversion table}\label{conversion}
To aid readers coming from various notational conventions we collect conversions of indices in binary, decimal and $\eta$ function expansion component notation.\\
\begin{center}
{\bf n=4}\\[2mm]
\begin{tabular}{ccc}
binary & decimal & $\eta$-function\\
\hline
$a_{0000}$&$a_{0}$&$F_{0}$\\
$a_{0001}$&$a_{1}$&$F_{4}$\\
$a_{0010}$&$a_{2}$&$F_{3}$\\
$a_{0011}$&$a_{3}$&$F_{34}$\\
$a_{0100}$&$a_{4}$&$F_{2}$\\
$a_{0101}$&$a_{5}$&$F_{24}$\\
$a_{0110}$&$a_{6}$&$F_{23}$\\
$a_{0111}$&$a_{7}$&$F_{234}$\\
$a_{1000}$&$a_{8}$&$F_{1}$\\
$a_{1001}$&$a_{9}$&$F_{14}$\\
$a_{1010}$&$a_{10}$&$F_{13}$\\
$a_{1011}$&$a_{11}$&$F_{134}$\\
$a_{1100}$&$a_{12}$&$F_{12}$\\
$a_{1101}$&$a_{13}$&$F_{124}$\\
$a_{1110}$&$a_{14}$&$F_{123}$\\
$a_{1111}$&$a_{15}$&$F_{1234}$\\
\end{tabular}
\end{center}
\appendix{Proof of factorization for n=2}\label{proof-w2}
When $F(\eta_1, \eta_2)$ factorizes, then obviously $w_{12}(F)=0$. Now, let
$F(\eta_1, \eta_2)=F_0+F_1\eta_1+F_2\eta_2+F_{12}\eta_1\eta_2$ and $w_{12}(F)=0$.
\begin{itemize}
\item Let $F_0, F_1, F_2 F_{12}\neq 0$, then $F_{12}=\frac{F_1F_2}{F_0}$.
Hence
\begin{equation}
F(\eta_1, \eta_2)=F_0\left(1+\frac{F_1}{F_0}\eta_1+\frac{F_2}{F_0}\eta_2+
\frac{F_1F_2}{F_0^2}\eta_1\eta_2\right)=
F_0\exp{\left\{\frac{F_1}{F_0}\eta_1+\frac{F_2}{F_0}\eta_2\right\}}
\end{equation}
and, e.g. $G(\eta_1)=F_0e^{\frac{F_1}{F_0}\eta_1}$ and $G(\eta_2)=e^{\frac{F_2}{F_0}\eta_2}$
\item Let $F_0\neq 0$, $F_{12}=0$, then e.g. $F_2=0$ and $F_1\neq 0$ and
$F(\eta_1, \eta_2)=1\cdot (F_0+F_1\eta_1)$
\item Let $F_0=0$, then $F_1$ or $F_2$ vanishes. Let $F_1\neq 0$
\begin{equation}
F(\eta_1, \eta_2)=F_1\eta_1+F_{12}\eta_1\eta_2=\eta_1(F_1+F_{12}\eta_2)
\end{equation}
Analogously for $F_2\neq 0$; $F=\eta_2(F_2+F_{12}\eta_1)$.
\end{itemize}
\appendix{Proof of the strong factorization criterion for n=3}\label{sft}
Let us consider the $(12)(3)$-separation. We have to show that
there exisits decomposition $F(\eta_1, \eta_2, \eta_3)=G(\eta_1, \eta_2)\tilde{G}(\eta_3)$ iff
\begin{eqnarray}
w_{13}(F)&=&0, \quad\quad w_{13}(\partial_2 F)=0,\\
w_{23}(F)&=&0,\quad\quad w_{23}(\partial_1 F)=0
\end{eqnarray}
Let $F=G(\eta_1, \eta_2)\tilde{G}(\eta_3)=G\tilde{G}$
\begin{equation}
w_{13}(F)=\left|
\begin{array}{cl}
G\cdot\tilde{G}&\partial_1G\cdot\tilde{G}\\
G\cdot\partial_3 \tilde{G}&\partial_1 G\cdot\partial_3 \tilde{G}
\end{array}
\right|=0
\end{equation}
and
\begin{equation}
w_{13}(\partial_2F)=\left|
\begin{array}{cl}
\partial_2G\cdot \tilde{G}&\partial_1\partial_2G\cdot\tilde{G}\\
\partial_2G\cdot\partial_3 \tilde{G}&\partial_1\partial_2 G\cdot\partial_3 \tilde{G}
\end{array}
\right|=0
\end{equation}
Analogously $w_{23}(F)=0=w_{23}(\partial_1 F)$.
Now, let us see how factorization follows from vanishing Wronskians. We have
\begin{eqnarray}
w_{13}(F|_{\eta_2=0})&=&0, \quad, \tilde{\mathcal{H}}_2=0, \quad w_{13}(\partial_2 F)=0,\\
w_{23}(F|_{\eta_1=0})&=&0, \quad, \tilde{\mathcal{H}}_1=0, \quad w_{13}(\partial_1 F)=0,\\
\end{eqnarray}
Taking $\frac{1}{2}(\tilde{\mathcal{H}}_1+\tilde{\mathcal{H}}_2)=0$ and
$\frac{1}{2}(\tilde{\mathcal{H}}_1-\tilde{\mathcal{H}}_2)=0$ we obtain the following set of conditions
\begin{eqnarray}
F_0F_{13}-F_1F_3&=&0,\\\label{w1}
F_0F_{23}-F_2F_3&=&0,\\\label{w2}
F_0F_{123}-F_3F_{12}&=&0,\\\label{w3}
F_1F_{23}-F_2F_{13}&=&0,\\\label{w4}
F_1F_{123}-F_{12}F_{13}&=&0,\\\label{w5}
F_2F_{123}-F_{12}F_{23}&=&0\label{w6}
\end{eqnarray}
For any $F(\eta_1, \eta_2, \eta_3)$ we can write
$
F(\eta_1, \eta_2, \eta_3)=\stackrel{(2)}{F}(\eta_1, \eta_3)+
\eta_2\stackrel{(2)}{\tilde{F}}(\eta_1, \eta_3)
$
with
\begin{equation}\label{f2}
\stackrel{(2)}{F}(\eta_1, \eta_3)=F_0+F_1\eta_1+F_3\eta_3+F_{13}\eta_1\eta_3=
F_0+F_1\eta_1+\eta_3(F_3+F_{13}\eta_1)
\end{equation}
and
\begin{equation}\label{f2t}
\stackrel{(2)}{\tilde{F}}(\eta_1, \eta_3)=F_2+F_{12}\eta_1+F_{23}\eta_3+F_{123}\eta_1\eta_3=
F_2+F_{12}\eta_1+\eta_3(F_{23}+F_{123}\eta_1)
\end{equation}
To find the factorization of the $F$ from conditions (\ref{w1}-\ref{w6}) we have to consider various forms of the $F$. The main families of solutions are labeled by $F_0=0$ (A) and $F_0\neq 0$ (B) and obtained as follows:\\
\underline{{\bf A:} $F_0\neq0$}.
\begin{enumerate}
\item $F_i,F_{ij}, F_{123}\neq0$. Using (\ref{w1}) and (\ref{w6}) above relations
(\ref{f2}) and (\ref{f2t}) can be written as
\begin{equation}
\stackrel{(2)}{F}=F_0e^{\frac{F_1}{F_0}\eta_1}\cdot e^{\frac{F_3}{F_0}\eta_3}
,\quad\quad
\stackrel{(2)}{\tilde{F}}=F_2e^{\frac{F_{12}}{F_2}\eta_1}\cdot e^{\frac{F_{23}}{F_2}\eta_3}
\end{equation}
and from (\ref{w2}) we get that, $e^{\frac{F_3}{F_0}\eta_3}=e^{\frac{F_{23}}{F_2}\eta_3}$.
\item $F_3=0$. Then $F_{13}=F_{23}=F_{123}=0$. Hence,
\begin{equation}
\stackrel{(2)}{F}=F_0e^{\frac{F_1}{F_0}\eta_1},
,\quad\quad
\stackrel{(2)}{\tilde{F}}=F_2e^{\frac{F_{12}}{F_2}\eta_1},
\end{equation}
therefore $\eta_3$ dependence factorizes trivially. Let us mention here that above reasoning in independent of the logarithm criterion, but (as discussed in the main text) is equivallent to the it.
\end{enumerate}
\underline{{\bf B:} $F_0=0$}.
\begin{enumerate}
\item $F_3=0$, $F_1=0$; and $F_2\neq0$, $F_{13}=0$. Here $\stackrel{(2)}{F}=0$ and $\stackrel{(2)}{\tilde{F}}$ has full expansion (\ref{f2t}), but $\eta_3$ dependence factorizes because of the Eq.(\ref{w6}).
\item $F_3=0$, $F_1=0$; and $F_2=0=F_{12}$. Hence,
\begin{equation}
\stackrel{(2)}{F}=F_{13}\eta_1\eta_3, \quad\quad
\stackrel{(2)}{\tilde{F}}=(F_{23}+F_{123}\eta_1)\eta_3
\end{equation}
\item $F_3=0$, $F_1=0$; and $F_2=0$, $F_{23}=0=F_{13}$. Here we have
\begin{equation}
\stackrel{(2)}{F}=0, \quad\quad
\stackrel{(2)}{\tilde{F}}=(F_{12}+F_{123}\eta_3)\eta_1
\end{equation}
%
%
\item $F_3=0$, $F_1\neq0$; and $F_{2}=0=F_{23}$. In this case
$\stackrel{(2)}{F}=F_1\eta_1e^{\frac{F_{13}}{F_1}\eta_3}$ and
$\stackrel{(2)}{\tilde{F}}=F_{12}\eta_1e^{\frac{F_{123}}{F_12}\eta_3}$. Due to (\ref{w5}) both functions have common factor $e^{\frac{F_{13}}{F_1}\eta_3}$.
\item $F_3=0$, $F_1\neq0$; and $F_{2}\neq0$, $F_{13}=F_{23}=F_{123}=0$. Hence
$\stackrel{(2)}{F}=F_1\eta_1$ and $\stackrel{(2)}{\tilde{F}}=F_2+F_{12}\eta_1$, thus factorization is trivial.
\item $F_3=0$, $F_1\neq0$; and $F_{2}, F_{12}, F_{13}, F_{23}, F_{123}\neq 0$. We have $\stackrel{(2)}{F}=F_1\eta_1e^{\frac{F_{13}}{F_1}\eta_3}$, then due to Eq. (\ref{w6}) $\stackrel{(2)}{\tilde{F}}=F_2e^{\frac{F_{12}}{F_2}\eta_1}
    e^{\frac{F_{23}}{F_2}\eta_3}$. Both functions have common factor $e^{\frac{F_{23}}{F_2}\eta_3}$, because of Eq. (\ref{w4}).
\item $F_3\neq 0$; and $F_1=F_2=F_{12}=0$. Here factorization is obvious,
$\stackrel{(2)}{F}=(F_3+F_{13}\eta_1)\eta_3$ and
$\stackrel{(2)}{\tilde{F}}=(F_{23}+F_{123}\eta_1)\eta_3$

\end{enumerate}
%
\appendix{Proof of the factorization condition of $F$ for n=4}\label{sf4}
\textbf{Case (i-j)(k-l):} Let $F(\eta_1, \eta_2,\eta_3,\eta_4)=G(\eta_i, \eta_j))\tilde{G}(\eta_k,\eta_l)$. We have
\begin{equation}
w_{ik}(F)=\left|
\begin{array}{cl}
G\cdot\tilde{G}&\partial_iG\cdot\tilde{G}\\
G\cdot\partial_k \tilde{G}&\partial_i G\cdot\partial_k \tilde{G}
\end{array}
\right|=0
\end{equation}
and also
\begin{equation}
w_{ik}(\partial_jF)=\left|
\begin{array}{cl}
\partial_jG\cdot \tilde{G}&\partial_i\partial_jG\cdot\tilde{G}\\
\partial_jG\cdot\partial_k \tilde{G}&\partial_i\partial_j G\cdot\partial_k \tilde{G}
\end{array}
\right|=0
\end{equation}
\begin{equation}
w_{ik}(\partial_j\partial_l F)=\left|
\begin{array}{cl}
\partial_jG\cdot \partial_l\tilde{G}&\partial_i\partial_jG\cdot\partial_l\tilde{G}\\
\partial_jG\cdot\partial_k\partial_l \tilde{G}&\partial_i\partial_j G\cdot\partial_k \partial_l\tilde{G}
\end{array}
\right|=0
\end{equation}

Analogously relevant determinants vanish for other configurations of indices.\\
\textbf{Case (i-j-k)(l):}
Let $F(\eta_1, \eta_2,\eta_3,\eta_4)=G(\eta_i, \eta_j,\eta_k)\tilde{G}(\eta_l)$ then,
\begin{equation}
w_{il}(F)=\left|
\begin{array}{cl}
G\tilde{G}&\partial_iG\cdot\tilde{G}\\
G\cdot\partial_l \tilde{G}&\partial_i G\cdot\partial_l \tilde{G}
\end{array}
\right|=0
\end{equation}
and also, as above,
\begin{equation}
w_{il}(\partial_j F)=\left|
\begin{array}{cl}
\partial_jG\cdot \tilde{G}&\partial_i\partial_jG\cdot\tilde{G}\\
\partial_jG\cdot\partial_l \tilde{G}&\partial_i\partial_j G\cdot\partial_l \tilde{G}
\end{array}
\right|=0
\end{equation}
and
\begin{equation}
w_{il}(\partial_j \partial_k F)=\left|
\begin{array}{cl}
\partial_j\partial_k G\cdot \tilde{G}&\partial_i\partial_j\partial_k G\cdot\tilde{G}\\
\partial_j\partial_k G\cdot\partial_l \tilde{G}&\partial_i\partial_j \partial_k G\cdot\partial_l \tilde{G}
\end{array}
\right|=0
\end{equation}
Analogous relations take place for other choices of indices $i$, $j$, $k$.
\appendix{n=3 Hyperdeterminant in terms of Wronskians}\label{Det-wron}
The hyperdeterminant is frequently written in the form
\begin{equation}\label{detf-w}
Det(F)=d_1-2d_2+4d_3,
\end{equation}
where
\begin{eqnarray}
d_1&=& F_0^2F_{123}^2+F_{3}^2F_{12}^2+F_{2}^2F_{13}^2+F_{1}^2F_{23}^2\\\nonumber
d_2&=& F_{0}F_{3}F_{12}F_{123}+F_{0}F_{2}F_{13}F_{123}+F_{0}F_{1}F_{23}F_{123}
+F_{2}F_{3}F_{13}F_{12}\\
&&+F_{1}F_{3}F_{23}F_{12}+F_{1}F_{2}F_{23}F_{13}\\
d_3&=&F_{0}F_{23}F_{13}F_{12}+F_{1}F_{2}F_{3}F_{123}.
\end{eqnarray}
Let us recall that
\begin{eqnarray}
\tilde{\mathcal{H}}_1&=&F_0F_{123}+F_1F_{23}-F_2F_{13}-F_3F_{12}\\
\tilde{\mathcal{H}}_2&=&F_0F_{123}-F_1F_{23}+F_2F_{13}-F_3F_{12}\\
\tilde{\mathcal{H}}_3&=&F_0F_{123}-F_1F_{23}-F_2F_{13}+F_3F_{12}
\end{eqnarray}
and
\begin{eqnarray}
w_{12}(F|_{\eta_3=0})&=& F_0F_{12}-F_1F_2, \quad\quad w_{12}(\partial_3 F)= F_3F_{123}-F_{13}F_{23}\\
w_{13}(F|_{\eta_2=0})&=& F_0F_{13}-F_1F_3, \quad\quad w_{13}(\partial_2 F)= F_2F_{123}-F_{12}F_{23}\\
w_{23}(F|_{\eta_1=0})&=& F_0F_{23}-F_2F_3, \quad\quad w_{23}(\partial_1 F)= F_1F_{123}-F_{12}F_{13}
\end{eqnarray}
%
Direct inspection shows that
\begin{equation}
\sum_{i=1}^3 \tilde{\mathcal{H}}_i=3d_1-2d_2
\end{equation}
and
\begin{equation}
\sum_{k=1}^3 w_{ij}(F|_{\eta_k=0})w_{ij}(\partial_k F) =d_2-3d_3
\end{equation}
Together with Eq. (\ref{detf-w}) we get relation (\ref{det-wron}). It is easy to get also analogous relation in the non-symmetric form e.g.
\begin{equation}
Det(F)=\tilde{\mathcal{H}}_1^2-4w_{23}(F|_{\eta_1=0})w_{23}(\partial_1F).
\end{equation}
\appendix{Representatives of entanglement SLOCC classes in naural form}\label{psi-4}
We present here the literal translation of states given by Chterental and Dokovi\'{c} in \cite{chte-dok}. Such representatives were firstly obtained by
Verstraete et al. in \cite{ver-deh-dem}, but in a slightly different form.
%
\begin{eqnarray}
\Psi_1&=&\frac{a+d}{2}e^{\vec{\eta}}+\frac{a-d}{2}(\eta_1\eta_2+\eta_3\eta_4)
+\frac{b+c}{2}(\eta_1\eta_3+\eta_2\eta_4)+\frac{b-c}{2}(\eta_1\eta_4+\eta_2\eta_3)
\\\nonumber
%
\Psi_2&=&\frac{a+c-i}{2}e^{\vec{\eta}}+\frac{a-c+i}{2}(\eta_1\eta_2+\eta_3\eta_4)
+\frac{b+c+i}{2}(\eta_1\eta_3+\eta_2\eta_4)\\
&&+\frac{b-c-i}{2}(\eta_1\eta_4+\eta_2\eta_3)
+\frac{i}{2}(\eta_1+ \eta_4+ \eta_2\eta_3\eta_4+\eta_1\eta_2\eta_3-\eta_2- \eta_3\\\nonumber
&&- \eta_1\eta_3\eta_4-\eta_1\eta_2\eta_4)
\\
%
%
\Psi_3&=&\frac{a}{2}e^{\eta_1\eta_2+\eta_3\eta_4}+\frac{b+1}{2}(\eta_1\eta_3+\eta_2\eta_4)
+\frac{b-1}{2}(\eta_1\eta_4+\eta_2\eta_3)
+\frac{1}{2}(\eta_3+\eta_1\eta_2\eta_4\\\nonumber
&&-\eta_4-\eta_1\eta_2\eta_3)
\\
%
%
\Psi_4&=&\frac{a+b}{2}e^{\vec{\eta}}+b(\eta_1\eta_3+\eta_2\eta_4)
+i(-\eta_2\eta_3+\eta_1\eta_4)+\frac{a-b}{2}(\eta_1\eta_2+\eta_3\eta_4)
\\\nonumber
&&+\frac{1}{2}(\eta_2+ \eta_3+ \eta_1\eta_3\eta_4+\eta_1\eta_2\eta_4-\eta_1- \eta_4- \eta_2\eta_3\eta_4-\eta_1\eta_2\eta_3)
\\
%
%
\Psi_5&=&\frac{a}{2}e^{\eta_1\eta_2+\eta_3\eta_4}
-2i(\eta_2+\eta_1\eta_4-\eta_1\eta_2\eta_3)
\\\nonumber
%
%
\Psi_6&=&\frac{a+i}{2}e^{\eta_1\eta_2+\eta_3\eta_4}+\frac{a+i+1}{2}(\eta_1\eta_3+\eta_2\eta_4)
+\frac{a-i-1}{2}(\eta_2\eta_3+\eta_1\eta_4)\\
&&+\frac{i+1}{2}(\eta_3+\eta_1\eta_2\eta_4)
+\frac{i-1}{2}(\eta_4+\eta_1\eta_2\eta_3)-
\frac{i}{2}(\eta_1+ \eta_2+\eta_2\eta_3\eta_4+\eta_1\eta_3\eta_4)
\\\nonumber
%
%
\Psi_7&=&\eta_1\eta_4+\eta_1\eta_3+\eta_2\eta_4-\eta_2\eta_3+\eta_1\eta_2
+\eta_1\eta_2\eta_3\eta_4+i(\eta_1\eta_4+\eta_1\eta_3-\eta_2-\eta_2\eta_3\eta_4\\
&&-\eta_1\eta_2\eta_4+\eta_1\eta_2\eta_3)
\\\nonumber
%
%
\Psi_8&=&\frac{i+1}{2}(e^{\vec{\eta}}-\eta_3-\eta_1\eta_2\eta_4)+
\frac{i-1}{2}(\eta_4+ \eta_1\eta_2\eta_3-\eta_3\eta_4-\eta_1\eta_2)
+\frac{1}{2}(\eta_2+\eta_1\eta_4+\eta_1\eta_3\\
&&+\eta_2\eta_3\eta_4+\eta_1+\eta_2\eta_3+\eta_2\eta_4
+\eta_1\eta_3\eta_4)-i(\eta_1+\eta_2\eta_4+\eta_2\eta_3+\eta_1\eta_3\eta_4)
\\
%
%
\Psi_9&=&\frac{1}{2}e^{\eta_1}(e^{\eta_2\eta_3\eta_4}+\eta_2+\eta_3\eta_4+i
(\eta_3+ \eta_4-\eta_2\eta_4-\eta_2\eta_3))
\end{eqnarray}
%
%
%

\begin{thebibliography}{99}
\bibitem{note1} This formula is implicit in considerations in Ref. \cite{staples},
where only the last term of the Equation (\ref{Stirling}) was important.
\end{thebibliography}

\label{lastpage}
\end{document}